\documentclass[12pt, a4paper]{article}%%%%where rsproca is the template name

\usepackage{amssymb,amsmath,mathrsfs,subfigure,epsfig,epstopdf}
\usepackage{amsmath}

\usepackage[latin1]{inputenc}

\usepackage{hyperref}
\usepackage{cleveref}
\usepackage{autonum}

\usepackage{geometry} % to change the page dimensions

\begin{document}
\setlength{\fboxsep}{1pt}
%%%% Article title to be placed here
\title{Wrinkles and creases in the bending, unbending and eversion of soft sectors}

\author{%%%% Author details
Taisiya Sigaeva$^{1}$, Robert Mangan$^{2}$,\\ Luigi Vergori$^{3}$, Michel Destrade$^{2}$ \\ and Les Sudak$^{1}$\\[12pt]
%%%%%%%%% Insert author address here
$^{1}$Department of Mechanical and Manufacturing Engineering, \\
University of Calgary, Calgary, AB, Canada;\\[6pt]
$^{2}$School of Mathematics, Statistics and Applied Mathematics,  \\
NUI Galway, University Road Galway, Ireland; \\[6pt]
$^{3}$Dipartimento di Ingegneria, \\ Universit\`{a} degli studi di Perugia, 06125 Perugia, Italy.}

\date{}

\maketitle

%%%% Abstract text to be placed here %%%%%%%%%%%%
\begin{abstract}
We study what is clearly one of the most common modes of deformation found in nature, science and engineering, namely the large elastic bending of curved structures, as well as its inverse, unbending, which can be brought beyond complete straightening to turn into eversion. We find that the suggested mathematical solution to these problems always exists and is unique when the solid is modelled as a homogeneous, isotropic, incompressible hyperelastic material with a strain-energy satisfying the strong ellipticity condition. We also provide explicit a\-symp\-to-tic solutions for thin sectors. 
When the deformations are severe enough, the compressed side of the elastic material may buckle and wrinkles could then develop.
We analyse in detail the onset of this instability for the Mooney-Rivlin strain energy, which covers the cases of the neo-Hookean model in exact non-linear elasticity and of third-order elastic materials in weakly non-linear elasticity. In particular the associated theoretical and numerical treatment allows us to predict the number and wavelength of the wrinkles. Guided by experimental observations we finally look at the development of creases, which we simulate through advanced finite element computations. In some cases the linearised analysis allows us to predict correctly the number and the wavelength of the creases, which turn out to occur only a few percent of strain earlier than the wrinkles.
\end{abstract}
%%%%%%%%%%%%%%%%%%%%%%%%%%%

%%%% Keyword entries to be placed here %%%%
\noindent
\textbf{keywords:} non-linear elasticity, instability, bending, unbending, eversion, cylindrical sector

%%%%%%%%%%%%%%

\section{Introduction}

%%%%%%%%%%%%%%%

Bending and unbending are without a doubt the two most common modes of deformation for the elastic curved structures found in nature and engineering. Mathematically, \emph{large} bending and unbending are actually exact solutions for incompressible, isotropic, non-linearly elastic circular sectors, as shown by Rivlin \cite{Rivl48}. 
Over the years there have been a good number of studies investigating the bending of a rectangular block into a sector of circular cylinder up to, and including the possible appearance of  wrinkles on the inner face of the resulting sector \cite{Haug99, CoDe08, DeAC09, RoGB10,Rudykh}. 
Very few works have looked at the stability of the converse problem, the straightening of a sector into a rectangular block  \cite{ DOSV14a, DOSV14b}, or at the stability of the bending into a closed full cylinder \cite{d-o-m}. 
The questions of existence, uniqueness and stability for the \emph{continuous} problem of bending and large unbending that can go all the way to eversion (when the inner and outer faces swap roles) remain scarcely investigated (only a few studies related to the deformation itself exist \cite{Horg05,Hill01,Sig16,Sig17}).
\begin{figure}[h]
\begin{center}
\includegraphics[width=0.9\linewidth]{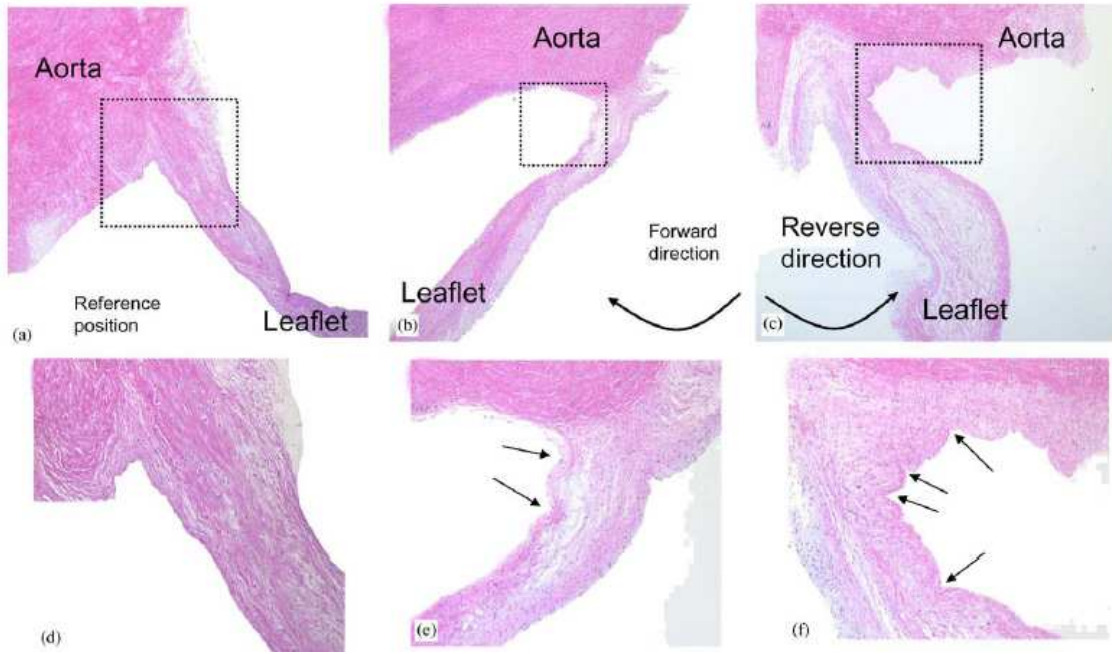}
\end{center}
\caption{Commissural region of an aortic valve leaflet (porcine heart): undeformed (left), bent during valve opening (middle), and unbent during valve closure (right).
The bottom pictures are the zooms indicated by the dotted squares in the top pictures. They show that wrinkles develop in both modes, eventually evolving into creases. 
Reprinted with permission from \cite{sacks06}.}
\label{leaflet}
\end{figure} 

Many works looking at large bending take their motivation from biological applications.  
An example can be found in the recent work by Rudykh and Boyce \cite{Rudykh} on the super flexibility of elasmoid fish in bending, due to the multilayered structure of their imbricated scale tissue. 
Similarly, researchers wishing to model residual stresses in tubular soft tissues often use the so-called ``opening angle method'', where the bending of a cylindrical sector into an intact tube creates large residual stresses -- see, for instance, the textbook by Taber \cite{Tabe04} for the modelling of residual stresses in arteries, in the left ventricle and in the embryonic heart. 

Wrinkles, in turn,  signal the onset of instability, and are often precursors to the development of creases, which are ubiquitous in nature, see the deformation of a heart valve leaflet in \ref{leaflet}, or the deep creases developed on the inner face of a depressurised pulmonary artery \cite{fung}.
These latter creases would considerably alter the blood flow during a low pressure episode due to an upstream blockage and alter the geometry of an artery for a planned surgery. 
In order to model  creases, we must first discover when the sector buckles on its way to be closed into a full cylinder.
With this ultimate goal in mind, we now embark on a complete resolution of the titular problem.

In the next section we recall the exact solution of non-linear elasticity for the flexure of circular sectors made of incompressible isotropic solids. 
We pay particular attention to the unbending mode, because it can be brought to go beyond the stage where the sector is deformed into a straight rectangular block. Then the sector becomes an everted sector and the inner and outer faces exchange roles. 

In \ref{Existence} we present analytical results for the existence and  uniqueness of the deformation. It turns out that bending, unbending and eversion of a cylindrical sector are always possible (and the solution is unique), provided its strain energy function satisfies the strong ellipticity condition. We also manage to provide an explicit thin-wall expansion of the results, valid for all strain energy functions up to the third order in the thickness. 
Details of the associated calculations are given in Appendix A.

In \ref{Wrinkles} and Appendix B we summarise our strategy to write down and solve numerically the boundary value problem of small-amplitude wrinkles superimposed on large bending, unbending or eversion, leaving the curved faces free of incremental traction. 
Within the framework of incremental elasticity \cite{ray}, we formulate the governing equations and the boundary conditions using the Stroh formalism. 
This formulation allows us to implement robust numerical procedures (surface impedance matrix method, compound matrix method) to overcome the numerical stiffness arising here. 

In \ref{Numerical results} we present the results of those numerical procedures for sectors made of Mooney-Rivlin materials or equivalently, of weakly non-linear, third-order elastic solids. The results turn out to be independent of the material constants, and are thus universal to these families of models. 
Our analysis of the number of wrinkles forming at the onset of instability is quite detailed and is consistent with, and thus generalises,  the previously studied special cases of bending of a rectangular block into a sector, unbending of a sector into a rectangular block, and bending of a sector into a full cylinder. 

In \ref{Table-top and Finite Element experiments} we present the results of table-top and finite element (FE) experiments of bending, unbending and everting a cylindrical sector. Both types of experiments reveal the formation of creases rather than sinusoidal wrinkles, in line with previous results for deforming homogeneous solids. 
In both cases, we  get period-doubling due to the merging of  some creases. 
The FE simulations show that the creases appear a bit earlier (a couple of percent less strain) than the wrinkles, which are thus not expressed. 
Nevertheless, the wrinkles analysis still proves useful, because we find that the number and the wavelength of the creases (counting the creases which would exist in the absence of period-doubling) predicted by the FE simulations is the same or close to the number and wavelength of wrinkles predicted by the numerical procedures of \ref{Numerical results}. 
It follows that the linearised analysis can be used to approximate the more computationally expensive FE simulations of creases, by predicting within a few percents the bifurcation strain, and the number and  wavelength of creases. 
It also generates the best shape possible for the perturbation introduced in the numerical creasing analysis.
Finally it forms the basis for the study of the stability of coated sectors, for which sinusoidal wrinkles are the dominant mode. 

%%%%%%%%%%%%%

\section{Large bending, unbending and eversion}
\label{Large bending, unbending and eversion}

%%%%%%%%%%%%%

We consider a right cylinder sector, initially undeformed and placed in the following region,
\begin{equation}
A\leq R\leq B, \qquad  -\alpha_r \le \Theta \le \alpha_r, \qquad \quad 0 \leq Z \leq L,
\end{equation}
where ($R,\Theta,Z$) are the cylindrical coordinates in the reference configuration, with orthonormal basis ($\mathbf E_R, \mathbf E_\Theta, \mathbf E_Z$).
Here $A$, $B$ are the inner and outer radii of the undeformed sector, respectively,  $L$ its axial length, and $2\alpha_r$ its \emph{undeformed} or \emph{referential angle}, related to its \emph{opening angle} $\alpha_o$ through the relation $\alpha_o=2(\pi-\alpha_r)$ . 

By applying appropriate moments and forces (determined later), the sector can be deformed into a more closed (bending) or more open (unbending) sector, with current axial length $\ell$ and \emph{deformation angle} $2\alpha_d$.

Hereon we exclude the possibility of scenarios when $\alpha_r$ and $\alpha_d$ are exact zeros, which are to be treated separately using Cartesian coordinate system and different universal solutions as in \cite{DOSV14a, CoDe08 }, for instance. Case $\alpha_r=0$ describes bending of a rectangular block, while case $\alpha_d=0$ corresponds to the problem of the sector straightening. 
Still, we encompass these cases in the limits where $\alpha_r \ll 1$ and $|\alpha_d| \ll 1$.
Hence we have
\begin{equation}
\alpha_r\in(0,\pi],\qquad \alpha_d\in[-\pi,\pi]-\{0\}.
\end{equation}

\begin{figure}[!h]
\centering\includegraphics[width=0.98\linewidth]{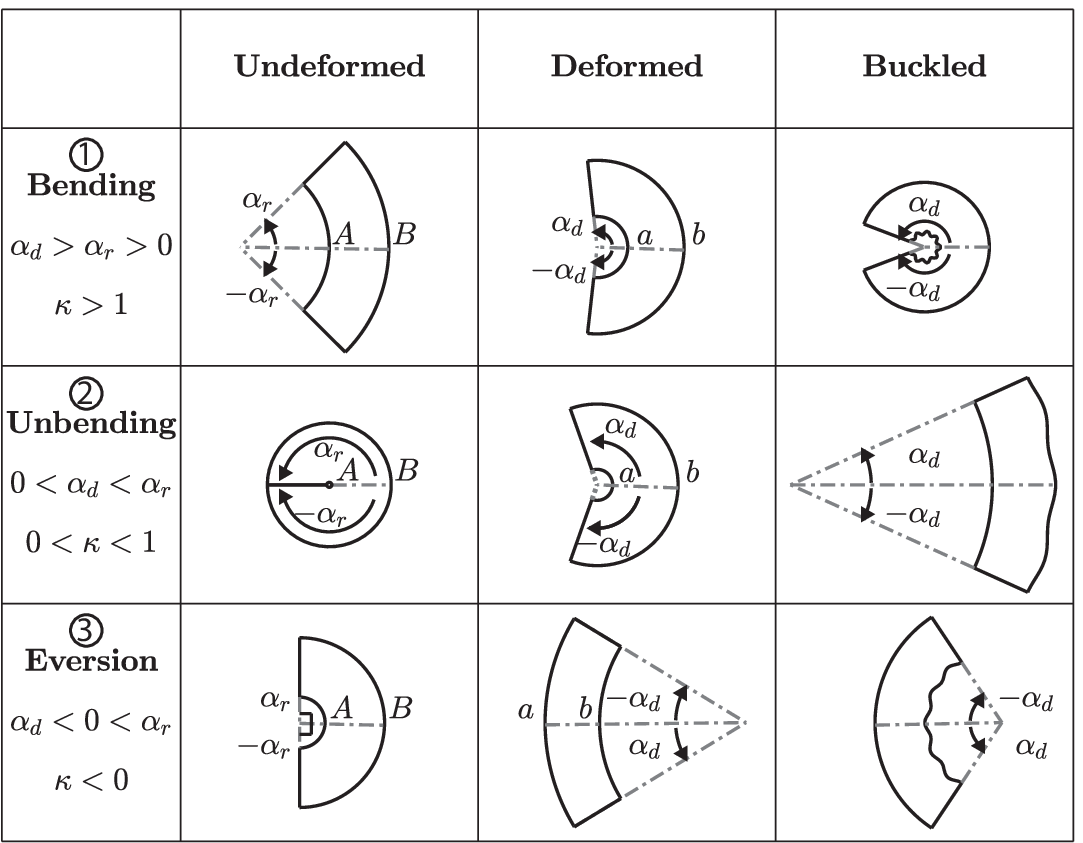}
\caption{The three scenarios covered in this paper for a cylindrical circular sector (second column), which is bent, unbent, or everted (third column) and eventually buckles on one face (last column).}
\label{sketches}
\end{figure}

We now introduce $\kappa$, a measure of the \emph{change in the angles}, as
\begin{equation}
\kappa = \dfrac{\alpha_d}{\alpha_r}\in\left[-\frac{\pi}{\alpha_r},\frac{\pi}{\alpha_r}\right]-\{0\}.
\end{equation}
Hence $\kappa >1$ corresponds to \emph{bending}, $\kappa <1$ corresponds to \emph{unbending}, and further, $\kappa <0$ corresponds to unbending beyond the straight rectangular configuration, a deformation which we call \emph{eversion} from now on. 
 \ref{sketches} shows sketches of these deformations and where wrinkling is going to take place.

%\begin{figure}[!h]
%\centering\includegraphics[width=1.0\linewidth]{fig2-draft.pdf}
%\caption{The three scenarios covered in this paper for a cylindrical circular sector (second column), which is bent, unbent, or everted (third column) and eventually buckles on one face (last column).}
%\label{sketches}
%\end{figure}

The deformation  can be modelled as
\begin{equation}\label{deformation1}
r=r(R),\qquad \theta=\kappa\Theta, \qquad z=\lambda_z Z,
\end{equation}
where $\lambda_z=\ell/L$ is the axial stretch, and ($r,\theta,z$) are the cylindrical coordinates in the current configuration, with orthonormal basis ($\mathbf e_r, \mathbf e_\theta, \mathbf e_z$).

We define the current radii $a \equiv r(A)$ and $b \equiv r(B)$. 
When $\kappa>0$, the inner and outer faces remain the respective inner and outer faces of the deformed sector, which occupies the following region,
\begin{equation}
a \le r \le b, \qquad -\alpha_d \le \theta \le \alpha_d, \qquad 0\le z \le \ell.
\end{equation}
But when $\kappa<0$ (eversion), the  inner face of the undeformed sector  becomes the outer face of the deformed sector, and \emph{vice versa}.
The sector then occupies the  region
\begin{equation}
b \le r \le a, \qquad \alpha_d \le \theta \le  -\alpha_d, \qquad 0\le z \le \ell.
\end{equation}
Note that the deformation does not account for slanting surfaces that can appear in large bending, especially in eversion \cite{Hill01}.

The corresponding deformation gradient $\mathbf F$ has components
\begin{equation}\label{deformation gradients}
\mathbf{F} = \dfrac{\text d r}{\text d R}\mathbf e_r \otimes \mathbf E_r + \dfrac{\kappa r}{R}\mathbf{e}_\theta\otimes\mathbf{E}_\Theta+\lambda_z\mathbf{e}_z\otimes\mathbf{E}_Z.
\end{equation}
For incompressible solids, $\det \mathbf{F}=1$ at all times, from which we deduce that
\begin{equation}\label{radii}
r = \sqrt{\frac{R^2-A^2}{\kappa\lambda_z}+a^2},
\qquad 
b=\sqrt{\frac{B^2-A^2}{\kappa\lambda_z}+a^2}.
\end{equation} 

Then we find the principal stretches (the square roots of the eigenvalues of $\mathbf{F}\mathbf{F}^T$) as
\begin{equation}\label{stretches}
\lambda_r = \dfrac{R}{|\kappa|\lambda_zr}, \qquad \lambda_\theta=\frac{|\kappa| r}{R},\qquad \lambda_z.
\end{equation}

Notice that in the special case where $\kappa \lambda_z = A^2/a^2$, the deformation is homogeneous. 
It then reads 
\begin{equation}\label{homogeneous}
r=\frac{R}{\sqrt{\kappa\lambda_z}}, \qquad \theta=\kappa\Theta, \qquad z=\lambda_z Z, 
\end{equation}
with constant principal stretches,
\begin{equation}\label{lambdas}
\lambda_r=\frac{1}{\sqrt{\kappa\lambda_z}},\qquad \lambda_\theta=\sqrt{\frac{\kappa}{\lambda_z}}, \qquad \lambda_z.
\end{equation}

Now we compute the forces and moments required to effect the deformation. 
For an incompressible, isotropic and hyperelastic material with strain energy  density $W=W(\lambda_r,\lambda_\theta,\lambda_z)$, the Cauchy stress tensor $\boldsymbol{\sigma}$ has components
\begin{equation}
\boldsymbol{\sigma} = \sigma_{rr}\mathbf e_r \otimes \mathbf e_r + \sigma_{\theta\theta}\mathbf{e}_\theta\otimes \mathbf{e}_\theta + \sigma_{zz}\mathbf{e}_z\otimes \mathbf{e}_z,
\qquad
\sigma_{qq}=-p+\lambda_q\frac{\partial W}{\partial \lambda_q}\qquad  ( q=r,\theta,z),
\end{equation}
where $p$ is the Lagrange multiplier introduced by the constraint of incompressibility.
Because the principal stretches  do not depend on $\theta$ and $z$,  we readily deduce from the equilibrium equations that $p=p(r)$ only, and that  
\begin{equation}\label{equilibrium}
\dfrac{\text d \sigma_{rr}}{\text d r} + \frac{\sigma_{rr}-\sigma_{\theta\theta}}{r} = 0,
\end{equation}
which must be solved subject to the boundary conditions of traction-free inner and outer faces:
\begin{equation}\label{boundary conditions}
\sigma_{rr}(a)=\sigma_{rr}(b)=0.
\end{equation}

To non-dimensionalise the equations, we use the \emph{scaled circumferential stretch} $\lambda$ (and its values $\lambda_a$, $\lambda_b$ on the faces), and the \emph{radii ratio} $\rho$, defined as 
\begin{equation}\label{not}
\lambda=\dfrac{\sqrt{\lambda_z}|\kappa|r}{R},\qquad  
\lambda_{a}=\dfrac{\sqrt{\lambda_z}|\kappa|a}{A},\qquad 
\lambda_{b}=\dfrac{\sqrt{\lambda_z}|\kappa|b}{B}, \qquad 
\rho=\frac{A}{B}\in]0,1[.
\end{equation}
According to \ref{radii} they are linked as follows, 
\begin{equation}\label{chib}
\lambda_{b}=\sqrt{\rho^2\lambda_{a}^2+(1-\rho^2)\kappa}.
\end{equation}
%Concerning the ordering of $\lambda_a$, $\lambda$ and $\lambda_{b}$, we distinguish the following cases
%\begin{itemize}
%\item[(\emph{a})] if $\kappa\in]0,2\pi/\alpha_r]$ and 
%\begin{itemize}
%\item[(\emph{i})] $0<\lambda_{a}<\sqrt{\kappa}$, then $\lambda_{a}\leq\lambda\leq\lambda_b<\sqrt{\kappa}$,
%\item[(\emph{ii})] $\lambda_{a}>\sqrt{\kappa}$, then $\sqrt{\kappa}<\lambda_{b}\leq\lambda\leq\lambda_a$;
%\item[(\emph{iii})] $\lambda_a=\sqrt{\kappa}$, then $\lambda_a=\lambda=\lambda_b$;
%\end{itemize}
%\item[(\emph{b})] if $\kappa\in[-2\pi/\alpha_r,0[$, then $\lambda_{b}\leq\lambda\leq\lambda_a$, and, in virtue of \ref{a}, $\lambda_a^2>(1-\rho^2)|\kappa|/\rho^2$.
%\end{itemize}

%\begin{equation}
%\lambda_1=\sqrt{\frac{1}{\lambda_z|\kappa|\chi}},\quad \lambda_2=\sqrt{\frac{|\kappa|\chi}{\lambda_z}}
%\end{equation}

Now we implement the change of variables from $r$ to $\lambda$ through 
\begin{equation}\label{change}
r\frac{\text d \lambda}{\text d r}=\frac\lambda\kappa(\kappa-\lambda^2),
\qquad \text{ or, equivalently, } \qquad
\text d r=\frac{a\lambda_a\kappa}{\kappa-\lambda^2}\sqrt{\frac{\kappa-\lambda_a^2}{\kappa-\lambda^2}}\text d\lambda.
\end{equation}
Then, introducing the single variable strain energy function $\widehat W = \widehat W(\lambda)$ as 
\begin{equation}\label{def w}
\widehat{W}(\lambda)\equiv W\left(1/(\sqrt{\lambda_z}\lambda),\lambda/\sqrt{\lambda_z},\lambda_z\right),
\end{equation}
we deduce that
\begin{equation}\label{difference}
\lambda\widehat{W}'(\lambda)=-\left(\lambda_1\frac{\partial W}{\partial\lambda_1}-\lambda_2\frac{\partial W}{\partial\lambda_2}\right)=-(\sigma_{rr}-\sigma_{\theta\theta}),
\end{equation}
so that the governing equation (\ref{equilibrium}) and the  boundary conditions (\ref{boundary conditions}) become, respectively,
\begin{equation}\label{equilibrium 2}
\dfrac{\text d\sigma_{rr}}{\text d\lambda} = \kappa\dfrac{\widehat{W}'(\lambda)}{\kappa-\lambda^2}, \qquad 
\sigma_{rr}(\lambda_{a})=0, \qquad \sigma_{rr}(\lambda_{b})=0.
\end{equation}
These can be integrated to give
\begin{equation}\label{sigma1}
\sigma_{rr}=\kappa\int_{\lambda_a}^{\lambda}\dfrac{\widehat{W}'(s)}{\kappa-s^2} \text d s,
\qquad
\int_{\lambda_{a}}^{\lambda_b}\frac{\widehat{W}'(s)}{\kappa-s^2} \text d s=0.
\end{equation} 
Notice that the latter equation and \ref{chib} form a system of two equations for $\lambda_a$, $\lambda_b$
(of course, it must be checked first that the equation has a solution, see next section).
Hence, if a given material is prescribed by the choice of its strain energy $W$, and the original dimensions $A$, $B$, $\alpha_r$ are prescribed, and the target deformation angle $\alpha_d$ is prescribed, then $\lambda_a$, $\lambda_b$ are found from these two equations, and the new radii $a$, $b$ follow.
 
Now that the radial stress $\sigma_{rr}$ is determined, we deduce the circumferential stress from \ref{difference} as
%\begin{equation}\label{azimutal}
$\sigma_{\theta\theta}=\sigma_{rr}+\lambda\widehat{W}'(\lambda)$.
%\end{equation}
Finally, we find that the stresses on the end surfaces $\theta=\pm \alpha_d$ are equivalent to couples with moments 
%\begin{equation}
%r=a\lambda_a\lambda\sqrt{\frac{\kappa-\lambda_a^2}{\kappa-\lambda^2}}
%\end{equation}
\begin{equation} 
\mathbf{M}_{|\theta=\pm\alpha_d}
= \pm\left\{A^2L\lambda_a^4(\kappa-\lambda_{a}^2)\int_{\lambda_a}^{\lambda_b}\frac{\widehat{W}'(s)}{(\kappa-s^2)^2}\text d s\right\}\mathbf e_z.
\end{equation}

%%%%%%%%%%%%%%%%%%%%%%%

\section{Existence, uniqueness and thin-wall expansion}
\label{Existence}

%%%%%%%%%%%%%%%%%%%%%%

We  investigated the existence and uniqueness of a positive root to \ref{sigma1}$_2$, and found that they are always guaranteed for materials with a  strain-energy function $W$ satisfying the \emph{strong ellipticity condition}. 
This condition simply puts constraints on the material parameters of many widely used models.
For example it is satisfied by the neo-Hookean, Mooney-Rivlin, Fung, Gent, and one-term Ogden models, as long as all parameters are positive \cite{DOSV14a}.
We relegate the details of this proof to Appendix A.

For \emph{thin sectors}, we were also able to establish some general conclusions about the deformed configuration.
For our asymptotic analysis we introduced the  following \emph{small thickness parameter} $\varepsilon>0$ defined as
\begin{equation}
\varepsilon= 1-\rho = (B - A)/B  \ll1.\label{epsi}
\end{equation}
Then we found the following expansion of $\lambda_a$ up to order $\varepsilon^4$:
\begin{multline}\label{thinexp}
 \lambda_a = 1+\frac{1}{2}(1-\kappa)\varepsilon 
 + \dfrac{1}{24}(1-\kappa)(13-3\kappa)\varepsilon^2-\frac{1}{48}(1-\kappa)(3\kappa^2+8\kappa-27)\varepsilon^3\\
+\frac{1}{5760}(1-\kappa)\Big[45\kappa^3-363\kappa^2-1813\kappa+3667\\
+(1-\kappa)^2\frac{2(15\kappa-23)\widehat{W}^{\mathrm{iv}}(1)-3(1-\kappa)\widehat{W}^{\mathrm{v}}(1)}{\widehat{W}''(1)}\Big]\varepsilon^4 + \mathcal O(\varepsilon^5).
\end{multline}
In particular, note that the results are independent of the form of strain-energy function up to order $\varepsilon^3$. 
Again the details are collected in Appendix A.

%%%%%%%%%%%%%%%%%%%%%%%

\section{Wrinkles}
\label{Wrinkles}

%%%%%%%%%%%%%%%%%%%%%%%

Incremental instability  is triggered by the apparition of \emph{small-amplitude wrinkles} on the compressed face of the deformed sector. 
For bending ($\kappa>1$) and eversion ($\kappa <0$), this is the inner face; for unbending with $0<\kappa<1$, it is the outer face, see the last column of  \ref{sketches}.  

The existence of small-amplitude wrinkles itself is governed by the incremental equations of incompressibility and of equilibrium.
These equations can be formatted into the so-called \emph{Stroh formulation}, a first-order system of linear equations with variable coefficients. 
We do not present the details of this derivation, which can be found in Destrade et al. \cite{d-o-m}. 

It suffices to recall that the incremental mechanical displacements $\mathbf u$ are sought in the form 
\begin{equation}\label{u components}
\mathbf u = \Re\{ [U(r)  \mathbf e_{r} + V(r) \mathbf e_\theta]\text e^{\text i n \theta} \},\qquad n=\frac{m\pi}{\alpha_d}=\frac{m\pi}{\kappa\alpha_r}\;(m\in\mathbb{N}),
\end{equation}
where the amplitudes $U$ and $V$ are functions of $r$ only, and $n$ is a real number to be determined from the condition of
no incremental normal tractions on the end faces $\theta=\pm\alpha_d$ of a sector; $m$ is 
 an integer, which we call the \emph{circumferential mode number}, giving the number of wrinkles on the contracted face.
Then the components of the incremental nominal traction $\dot{\mathbf S}$ have the same structure:
\begin{equation}
\dot{\mathbf S}^{\mathrm T} \mathbf e_{r} = \Re\{ [S_{rr}(r)  \mathbf e_{r} + S_{r\theta} (r) \mathbf e_\theta]\text e^{\text i n \theta}\}.
\end{equation}

We can readily obtain the equations for the displacement-traction Stroh vector $\boldsymbol{\eta}=[U,V,\mathrm{i} r S_{rr},\mathrm{i} rS_{r\theta}]^{\mathrm T}$ in the form  \cite{d-o-m}
\begin{equation}\label{stroh}
\dfrac{\text d}{\text d r}\boldsymbol{\eta}(r)=\frac{\mathrm{i}}{r}\mathbf{G}(r)\boldsymbol{\eta}(r),
\end{equation}
where $\mathbf G$ is the Stroh matrix:
\begin{equation}\label{gi}
\mathbf G=\left(\begin{array}{cccc}
\mathrm{i} & -n & 0 & 0\\
[3mm]
-n(1-\sigma_{rr}/\alpha) & -\mathrm{i}(1-\sigma_{rr}/\alpha) & 0 & -1/\alpha\\
[3mm]
\kappa_{11} & \mathrm{i}\kappa_{12} & -\mathrm{i} & -n(1-\sigma_{rr}/\alpha)\\
[3mm]
-\mathrm{i} \kappa_{12} & \kappa_{22} & -n & \mathrm{i}(1-\sigma_{rr}/\alpha)
\end{array}\right).
\end{equation}
Here 
\begin{equation}\label{coefficients}
\left.\begin{array}{ll}
 \alpha=\dfrac{\lambda\widehat{W}'(\lambda)}{\lambda^4-1},\qquad 
\gamma=\lambda^4\alpha, 
\qquad \beta=\dfrac{\lambda^2}{2}\widehat{W}''(\lambda)-\alpha, \\[6pt]
 \kappa_{11}=2(\alpha+\beta-\sigma_{rr})+n^2[\gamma-\alpha(1-\sigma_{rr}/\alpha)^2],  \\[6pt]
 \kappa_{12}=n[2\beta+\alpha+\gamma-{\sigma_{rr}}^2/\alpha], \\[6pt]
 \kappa_{22}=\gamma-\alpha(1-\sigma_{rr}/\alpha)^2+2n^2(\alpha+\beta-\sigma_{rr}).
\end{array}\right.
\end{equation}

Now the system (\ref{stroh}) needs to be integrated numerically, subject to the boundary conditions that the incremental traction vanish on the inner and outer faces, i.e.
\begin{equation}\label{bcstroh}
S_{rr}(a) =S_{rr}(b) = 0, \qquad S_{r\theta} (a)=S_{r\theta} (b)= 0.
\end{equation}

\section{Numerical results for wrinkles}
\label{Numerical results}

%%%%%%%%%%%%%%%%%%%%%%%%%%%%%%%

The  numerical techniques described in Appendix B can be implemented to predict the onset of instability  in sectors made of any hyperelastic material.
From now on we specialise our discussion to \emph{Mooney-Rivlin solids}, for which
\begin{equation}
W = \dfrac{C}{2}\left(\lambda_1^2 + \lambda_2^2 + \lambda_3^2 -3\right) + \dfrac{D}{2}\left(\lambda_1^2\lambda_2^2 + \lambda_2^2\lambda_3^2 + \lambda_3^2\lambda_1^2 - 3\right),
\end{equation}
where $C \ge 0$ and $D \ge 0$ are material constants. 

For this class of nonlinearly elastic materials, the single variable function $\widehat{W}$ reads
\begin{equation}\label{mrsv}
\widehat{W}(\lambda) = \dfrac12\left[\left(C\lambda_z^{-1} + D\lambda_z\right)(\lambda^2 + \lambda^{-2}) + C(\lambda_z^2 - 3) + D\left(\lambda_z^{-2} - 3\right)\right],
\end{equation}
from which the parameters $\alpha$, $\beta$ and $\gamma$ in \ref{coefficients},  and the radial stress $\sigma_{rr}$ are found to be
\begin{align}
\label{abc}
& \alpha= \left(C \lambda_z^{-1} + D\lambda_z\right) \lambda^{-2},\qquad 
\gamma=\left(C\lambda_z^{-1} + D\lambda_z\right)\lambda^2,\qquad 
\beta= (\alpha + \gamma)/2,
\\[6pt]
\label{stress1}
&\sigma_{rr} = -\left(C\lambda_z^{-1}+D\lambda_z\right)\left[\frac{\kappa^2-1}{2\kappa}\ln\left(\frac{\lambda^2-\kappa}{\lambda_a^2-\kappa}\right)+\frac{1}{\kappa}\ln\left(\frac{\lambda}{\lambda_a}\right)+\frac{\lambda^2-\lambda_a^2}{2\lambda^2\lambda_a^2}\right].
\end{align}
The scaled circumferential stretch  $\lambda_a$ in \ref{stress1} is  the unique root of the equation
\begin{equation}\label{eqmr}
\frac{\kappa^2-1}{\kappa}\ln\rho+\frac{1}{\kappa}\ln\frac{\sqrt{\rho^2\lambda_a^2+(1-\rho^2)\kappa}}{\lambda_a}+\frac{(1-\rho^2)(\kappa-\lambda_a^2)}{2\lambda_a^2[\rho^2\lambda_a^2+(1-\rho^2)\kappa]}=0,
\end{equation}
which is derived from \ref{chib} and \ref{sigma1}$_2$ for the Mooney-Rivlin $\widehat{W}$  of \ref{mrsv}.

It can be easily shown that, in agreement with the thin-wall expansion (\ref{thinexp}), the unique solution to \ref{eqmr} tends to 1 as $\rho\rightarrow1$ for any value of $\kappa$. 
For sectors with small radii ratio $\rho = A/B$ we distinguish the bending ($\kappa > 1$), unbending ($0<\kappa<1$) and eversion ($\kappa<0$) cases and obtain the following respective approximations for the scaled circumferential stretch on the side under compression (i.e. $\lambda_a$ in bending, $\lambda_b$ in unbending and eversion):
\begin{equation}\label{approx}
\left.\begin{array}{ll}
\lambda_a \simeq \sqrt{\dfrac{\kappa}{W_0(\mathrm{e}\rho^{2(1-\kappa^2)})}} \qquad & \text{ if } \kappa>1,\\[18pt]
\lambda_b \simeq \sqrt{\dfrac{\kappa}{1+W_0\left(-\displaystyle{\rho^{2\kappa^2}}/{\mathrm{e}}\right)}} \qquad &\textrm{ if }0<\kappa<1,\\[18pt]
\lambda_b \simeq \sqrt{\dfrac{\kappa}{1+W_{-1}\left(-\displaystyle{\rho^{2\kappa^2}}/{\mathrm{e}}\right)}} \qquad &\textrm{ if }\kappa<0,
\end{array}\right.
\end{equation}
where $W_0$ and $W_{-1}$ are, respectively,  the upper and lower branches of the real-valued Lambert-W function. From \ref{approx}$_1$ (resp., \ref{approx}$_3$) we deduce that for bending (respectively, eversion) the scaled circumferential stretch  $\lambda_a$  (respectively, $\lambda_b$)  is an infinitesimal quantity of the same order as $1/\sqrt{|\ln \rho|}$ when $\rho\rightarrow 0$ and thus it tends abruptly to 0 as $\rho\rightarrow 0$. 
For unbending, $\lambda_b\rightarrow\sqrt{\kappa}$ as $\rho\rightarrow 0$ and 
\begin{equation}
\displaystyle\frac{\text{d}\lambda_b}{\text{d} \rho}\rightarrow\left\{\begin{array}{ll}
+\infty \quad & \textrm{if } \kappa\in\left(0,\displaystyle\frac{1}{\sqrt{2}}\right),\\
[5mm]
\displaystyle\frac{1}{2\mathrm{e}\sqrt[4]{2}} \quad & \textrm{if } \kappa=\displaystyle\frac{1}{\sqrt{2}},\\
[5mm]
0 \quad & \textrm{if } \kappa\in\left(\displaystyle\frac{1}{\sqrt{2}},1\right).
\end{array}\right.
\end{equation}
Once again, for $\kappa\in(0,1/\sqrt{2})$ the stretch on the side under compression changes rapidly near $\rho=0$.
 Because of the asymptotic behaviour of the circumferential stretches $\lambda_a$ and $\lambda_b$, the stability problem for a sector with a very small $\rho$ is numerically stiff.

Now from \ref{abc} and \ref{stress1} we can readily compute all the coefficients (\ref{coefficients}) of the Stroh matrix (\ref{gi}). 
As shown by Destrade et al. \cite{d-o-m} and from \ref{abc}-\ref{stress1}, the incremental governing equations and boundary conditions can then be normalised in such a way that $C$, $D$ and $\lambda_z$ disappear (simply by dividing all equations across by $C\lambda_z^{-1}+D\lambda_z$). 
These quantities thus play no role in the stability analysis, and the following results are thus valid for all values of $C$, $D$, $\lambda_z$.
This flexibility makes the results quite general, because the Mooney-Rivlin model recovers not only the neo-Hookean model of exact non-linear elasticity ($D=0$) but also, at the same order of approximation, the general form of strain energy function for weakly non-linear third-order (isotropic, incompressible) elasticity \cite{RiSa51,Destrade2010b}: 
\begin{equation}\label{TO_Energy}
W=\mu \,\mathrm{tr}(\mathbf{E}^2)+\frac{A}{3}\,\mathrm{tr}(\mathbf{E}^3),
\end{equation}
where $\mu>0$ is the second-order Lam\'e coefficient, $A$ is the third-order Landau constant and $\mathbf{E}=(\mathbf{F}^\text{T}\mathbf F -\mathbf{I})/2$ is the Green-Lagrange strain tensor.

\begin{figure}[!ht]
\centering\includegraphics[width=1.0\linewidth]{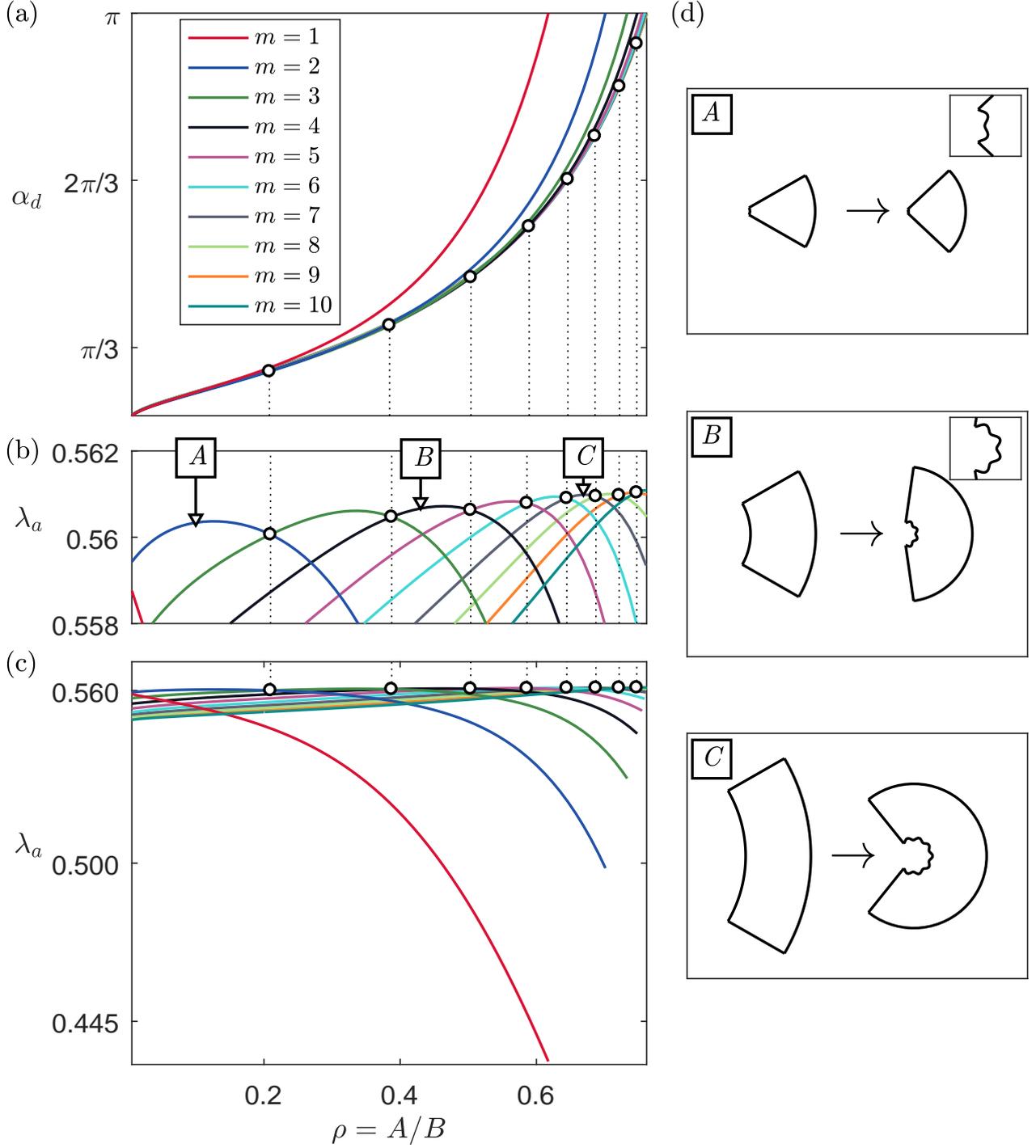}

\caption{Critical deformation angles $\alpha_d$ (a) and critical stretches $\lambda_a$ [(b): scaled and (c): regular] plotted versus radii ratios $\rho=A/B$ for sectors with $\alpha_r=\pi/6$ and mode numbers $m=1,\cdots,10$ undergoing plane strain bending. $(r,\theta)$ plane view (d) of the sectors indicated in (b) in the undeformed configuration and at buckling.
The numerical results for the specific sectors \fbox{$A$}, \fbox{$B$}, \fbox{$C$} are shown in \ref{tab:Table_1};
in (d), the lengths are normalised with respect to the initial thickness $H=B-A$, so that $A=\rho H/(1-\rho)$ and $B=H/(1-\rho)$.}
\label{fig_sim_1}
\end{figure}

\begin{figure}[!h]
\centering\includegraphics[width=1.0\linewidth]{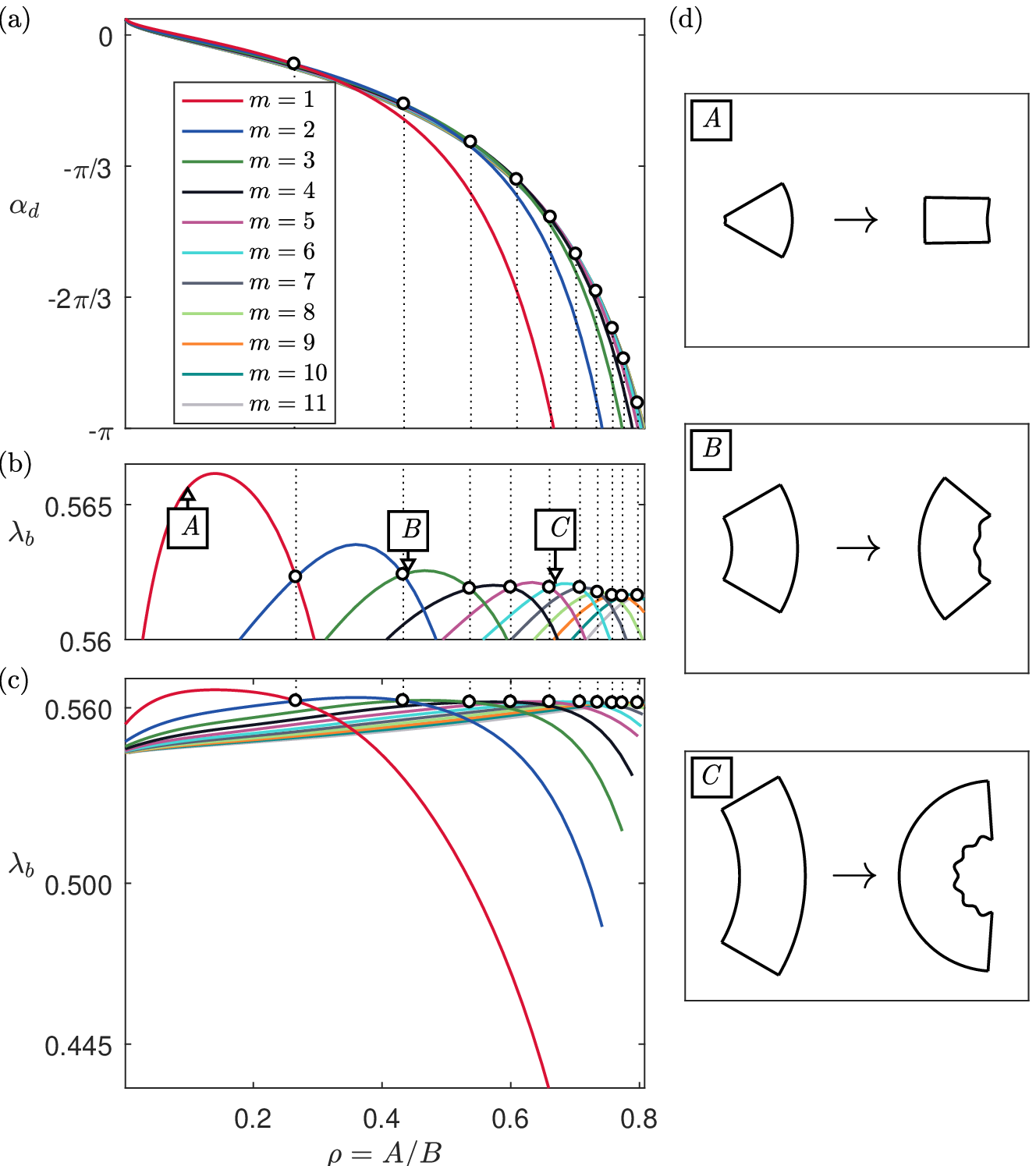}

\caption{Critical deformation angles $\alpha_d$ (a) and critical stretches $\lambda_b$ (b -- scaled and c -- regular) plotted versus radii ratios $\rho=A/B$ for sectors with $\alpha_r=\pi/6$ and mode numbers $m=1,\cdots,11$ undergoing plane strain unbending ($\alpha_r>\alpha_d>0$) and eversion ($\alpha_d<0$). $(r,\theta)$ plane view (d) of the sectors indicated in (b) in the undeformed configuration and at buckling.
The numerical results for the specific sectors \fbox{$A$} -- \fbox{$D$} are shown in  \ref{tab:Table_1}; in (d), the lengths are normalised with respect to the initial thickness $H=B-A$, so that $A=\rho H/(1-\rho)$ and $B=H/(1-\rho)$.}
\label{fig_sim_2}
\end{figure}

 \ref{fig_sim_1} and \ref{fig_sim_2} report the critical values of the bending angles $\alpha_d$ and the critical circumferential stretches on the corresponding contracted faces as functions of the radii ratio $\rho=A/B$ for a sector with an undeformed angle $\alpha_r=\pi/6$. 
In bending ( \ref{fig_sim_1}) the critical thresholds for $\alpha_d$ and $\lambda_a$ are plotted for $\rho\in(0,0.7619)$ because in this range, $\alpha_d>\pi$. 
As $\rho$ approaches 0.7619, $\alpha_d$ approaches $\pi$. 
Hence a circular cylindrical sector with $\rho\in(0.7619,1)$ can be closed to form an intact tube without experiencing wrinkles on the inner face $r=a$. 
In unbending/eversion ( \ref{fig_sim_2}) the critical thresholds are plotted for $\rho\in(0,0.8079)$. 
Here we see that a sector with $\rho\in(0.8079,1)$ can be completely everted to form an intact tube without the appearance of wrinkles on the inner side $r=b$.

 \ref{fig_sim_1} and \ref{fig_sim_2} display curves corresponding to different circumferential mode numbers $m=1, \ldots N$, which define the number of prismatic wrinkles appearing on the contracted side of a deformed sector. 
However, only the modes corresponding to the highest critical stretches $\lambda_a$ and $\lambda_b$ (correspondingly, the lowest critical angles $\alpha_d$ in bending and the highest critical angles $\alpha_d$ in unbending and eversion) are meaningful, as the lower stretches cannot be reached once a sector has buckled. 
We call these mode numbers the \textit{acute mode numbers}. 

For example,  \ref{fig_sim_1}a-c show that the acute mode number for a sector with $\alpha_r=\pi/6$ and $\rho\in(0,0.21)$ is $m=2$; 
for a sector with $\rho\in(0.21,0.39)$, it is $m=3$; and so on.
We use circle markers to highlight the transitions from one acute mode number to another as shown in \ref{fig_sim_1}a-c and \ref{fig_sim_2}a-c.

Now we provide a more in-depth examination of how the critical deformations and number of wrinkles in bending and eversion depend on the referential geometry; in particular, how they differ for the same sector in bending and eversion (unbending $\alpha_r>\alpha_d>0$ is not that noticeable in \ref{fig_sim_2}a for $\alpha_r=\pi/6$ and, thus, will be illustrated in the subsequent discussion for another $\alpha_r$). 
To this end, we pick the geometries labeled \fbox{$A$}, \fbox{$B$}, \fbox{$C$} as shown in \ref{fig_sim_1}b, \ref{fig_sim_2}b,  denoting sectors with initial radii ratios $\rho \simeq 0.1$, $0.486$, $0.67$, respectively, and referential angle $\alpha_r=\pi/6$. 
In \ref{fig_sim_1}d and \ref{fig_sim_2}d we show an  $(r,\theta)$ plane view of Sector \fbox{$A$} in the reference configuration and at buckling (the lengths are normalised with respect to the initial thickness $H=B-A$). 
 \ref{tab:Table_1} collects the results of the incremental stability analysis.
We note that  these sectors \fbox{$A$}, \fbox{$B$}, \fbox{$C$} present one more wrinkle in bending than in eversion, although this is not always the case for other referential geometries. 

\begin{table}
\centering
\begin{tabular}{lllllllll }
 &    &   & \multicolumn{3}{| l }{Bending}     & \multicolumn{3}{ | l }{Eversion}   \\
\hline
 Sector&$\rho = A/B$ & $\alpha_r$ & $m$ & $\alpha_d$ & $\lambda_a$  & $m$ & $\alpha_d$ & $\lambda_b$  \\
\hline
\fbox{$A$} & 0.1     &   $\pi/6$   & 2 & $0.2425\pi$   &0.5604  & 1 & $-0.0049\pi$ & 0.5654  \\ 
\fbox{$B$} & 0.486 &   $\pi/6$   & 4 & $0.4562\pi$ & 0.5607 & 3 & $-0.2204\pi$   & 0.5614  \\
\fbox{$C$} & 0.67   &   $\pi/6$   & 7 & $0.7175\pi$ & 0.5609 & 6 &  $-0.4802\pi$    & 0.5619 \\
\fbox{$D$} & 0.486   &   $\pi/3$   & 9 & $0.9124\pi$ & 0.5607 & 6 &  $-0.4372\pi$    & 0.5626   
\\\hline
\end{tabular}
\caption{Critical parameters for bending and eversion of three sectors from \ref{fig_sim_1}-\ref{fig_sim_2} and one sector from the following section.}
\label{tab:Table_1}
\end{table}%%%End of the table

To validate our results here, we connect with the analysis of Haughton \cite{Haug99} for the flexure of rectangular blocks. 
Even though we have an initial curvature for our sectors (as opposed to the rectangular geometry from \cite{Haug99}), we similarly discover that the different acute mode numbers effectively form an envelope  predicting loss of stability at the critical stretch of approximately $\lambda=0.563$ (\ref{fig_sim_1}c and \ref{fig_sim_2}c), see also \cite{DeAC09}.

\begin{figure}[!h]
\centering\includegraphics[width=1.0\linewidth]{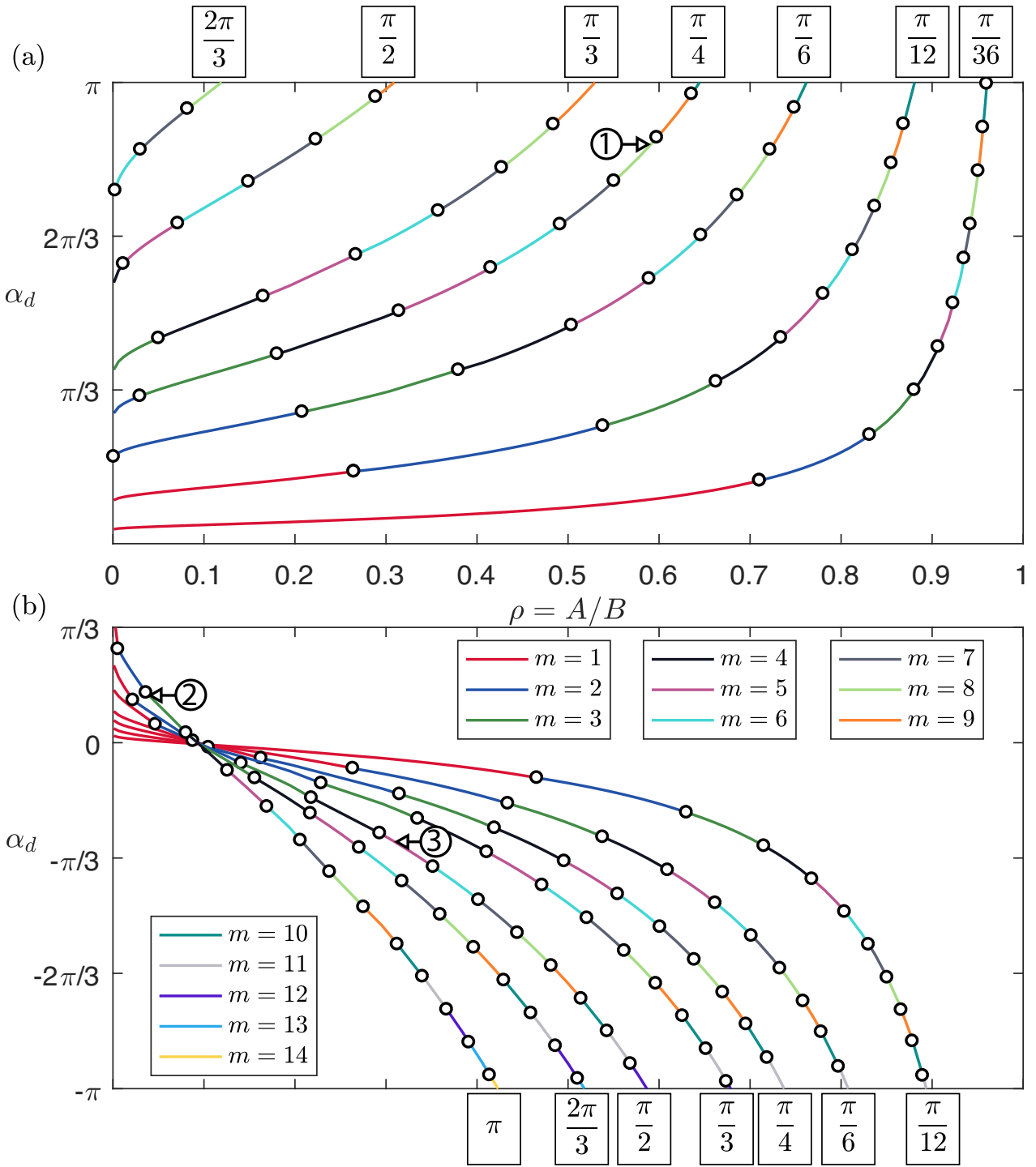}
%%% where xxxxxx name represents "figurename.eps"
\caption{Critical deformation angles $\alpha_d$ in bending (a), unbending (b, $\alpha_r>\alpha_d>0$) and eversion (b, $\alpha_d<0$) plotted versus radii ratios $\rho=A/B$ for sectors with different referential angles \fbox{$\alpha_r$} and acute mode numbers $m=1,\cdots,14$.
}
\label{fig_sim_3}
\end{figure}

The  discussion above was conducted for sectors with $\alpha_r=\pi/6$. 
In  \ref{fig_sim_3} we provide the critical deformation angles $\alpha_d$ as functions of the radii ratio $\rho=A/B$  for sectors corresponding to other relevant undeformed angles $\alpha_r$ (labeled by \fbox{$\displaystyle D$}). 
 \ref{fig_sim_3}a illustrates  critical deformations for bending and \ref{fig_sim_3}b for unbending ($\alpha_r>\alpha_d>0$) and eversion ($\alpha_d<0$). 
Only the curves corresponding to the acute mode numbers and points of transition between them are displayed, while the other information is not reported (compare, for illustration, critical deformation angles $\alpha_d$ for sectors with $\alpha_r=\pi/6$ in \ref{fig_sim_1}a, \ref{fig_sim_2}a and sectors with $\alpha_r=\pi/6$ in \ref{fig_sim_3}). 
Labels \textcircled{$1$}, \textcircled{$2$}, \textcircled{$3$} are used to indicate buckling states of sectors illustrated in \ref{sketches}.

For each referential angle $\alpha_r$, there exist limiting radii ratios  $\rho_{\alpha_r}^\star,\rho_{\alpha_r}^{\star\star}\in(0,1)$ such that $\alpha_d>\pi$ for all $\rho\in(\rho_{\alpha_r}^\star,1)$ in bending, and $\alpha_d<-\pi$ for all $\rho\in(\rho_{\alpha_r}^{\star\star},1)$ in eversion. 
This means that  a sector with $\rho\in(\rho^\star_{\alpha_r},1)$ (resp., $\rho\in(\rho^{\star\star}_{\alpha_r},1)$) can be closed (respectively, completely everted) to form an intact tube without the appearance of wrinkles on the inner side.  \ref{tab:Table_2} reports the radii ratios $\rho_{\alpha_r}^\star$ and $\rho_{\alpha_r}^{\star\star}$ for all the referential angles in \ref{fig_sim_3} as well as  the corresponding acute mode numbers  in bending and  eversion.

 \ref{fig_sim_3} and  \ref{tab:Table_2} contain then all the required information to form a theoretical prediction on whether, when and how a given sector will wrinkle.  
Notice that the overall largest acute number was found to be $m=14$, in eversion, as shown in \ref{fig_sim_3}b. 
For shorter wavelengths (larger $m$), the buckling occurs for sectors with deformed angles such that $|\alpha_d| >\pi$, which is physically impossible.

\begin{table}[!h]
	\centering
		\begin{tabular}{ccccc}
			 & \multicolumn{2}{| l }  {Bending}  & \multicolumn{2}{| l }  {Eversion} \\
			\hline
			$\alpha_r$ & $\rho_{\alpha_r}^\star$ & $m$ & $\rho_{\alpha_r}^{\star\star}$ & $m$\\
			\hline
			$\pi$  &  - & - & 0.423 & 14 \\
			$2\pi/3$ & 0.1193 & 8 & 0.518 & 13\\
			$\pi/2$ & 0.3092 &  9 & 0.5866 & 12\\
			$\pi/3$ & 0.5295 & 9 & 0.6787 & 12\\
			$\pi/4$ & 0.6445 & 10 & 0.7374 & 11\\
			$\pi/6$ & 0.7619 & 10 & 0.8079 & 11\\
			$\pi/12$ & 0.8808 & 10 & 0.8936 & 11\\
			$\pi/36$ & 0.9603 & 11 & - & -
			\\\hline
		\end{tabular}
		\caption{Values of the limiting radii ratios $\rho^\star_{\alpha_r}$ and $\rho_{\alpha_r}^{\star\star}$ for different values of the the referential angle $\alpha_r$. The acute mode numbers $m$ indicate the number of wrinkles appearing on the inner face of the intact tube obtained by bending (resp., everting) a sector with exact radii ratio $\rho_{\alpha_r}^\star$ (resp., $\rho_{\alpha_r}^{\star\star}$).
		Sectors with radii ratio greater than $\rho^\star_{\alpha_r}$ (resp., $\rho_{\alpha_r}^{\star\star}$) can be completely closed (everted) into an intact tube without wrinkling. 		}
\label{tab:Table_2}
\end{table}

Finally, to illustrate the generality of our wrinkling analysis, we make the connection with three special cases already reported in the literature: closing of a cylindrical sector into an intact tube \cite{d-o-m}, straightening of a cylindrical sector into a rectangular block \cite{DOSV14a, DOSV14b}, and bending of a rectangular block into a sector of a circular cylinder \cite{DeAC09}. 

In the case of closing of a sector into an intact tube on the onset of instability, we recover the critical deformations from \cite{d-o-m} by simply looking at the limiting values of $\alpha_d=\pi$ for different $\alpha_r$ (\ref{fig_sim_3}a). 
What is novel here is that our solution can address the situation when the sector buckles before the full closing (\ref{fig_sim_3}b, $\alpha_d<\pi$). 

Similar observations can be made for the straightening of a cylindrical sector. 
Critical deformations from \cite{DOSV14a} can be retrieved when we take $\alpha_d=0$ for different $\alpha_r$ (\ref{fig_sim_3}b) and again, our treatment is able to predict how sectors  buckle before the exact straightening (\ref{fig_sim_3}b, $0<\alpha_d<\alpha_r$). 

Next, when making the link with stability results for bending of a rectangular block, we are not able to derive the solution for $\alpha_r=0$ due to characteristic singularity, but we can simply consider a small referential angle $\alpha_r=\pi/36$, say \ref{fig_sim_3}a, and the corresponding critical deformations then match the results from \cite{DeAC09} very well.
 
To sum up, from the problem of bending, unbending and eversion of a cylindrical sector we are able to recover critical deformations of three classical universal deformations of incompressible non-linear elasticity. Moreover, it allows us to obtain new results on critical deformations of a circular sector bent, unbent or everted  (\ref{fig_sim_3}b, $\alpha_d<0$) into another sector.

%%%%%%%%%%%%%%%%%%%%%%%%%%%%

\section{Numerical results for creases}
%\section{Table-top and Finite Element experiments. \textcolor[rgb]{0,0,0}{Numerical results for creases?}}
\label{Table-top and Finite Element experiments}

%%%%%%%%%%%%%%%%%%%%%%%%%%%%%

\begin{figure}[h]
\begin{center}
\includegraphics[width=1.0\linewidth]{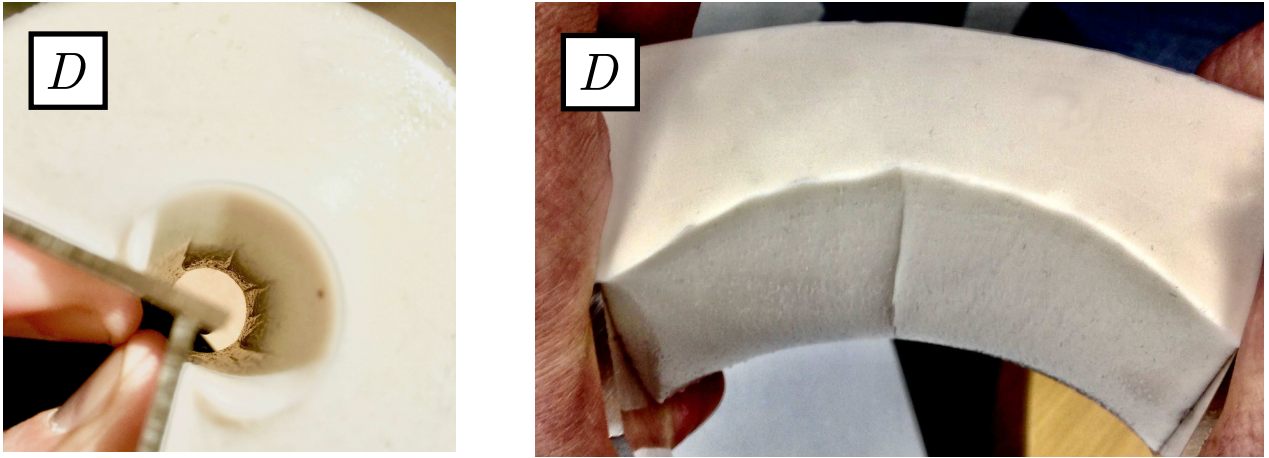}
\end{center}
\caption{Bending (left) and eversion (right) of a cylindrical sector made of silicone, with reference  angle $\alpha_r=\pi/3$ and radii ratio $\rho = A/B = 0.486$. Notice how creases appear on the contracted face, not wrinkles. We denote this physical sector as Sector \fbox{$D$}.  In bending, we count about six or seven creases; in eversion, only one.}
\label{silicone}
\end{figure}

Our treatment of small-amplitude wrinkles superimposed on large bending and unbending presented in the previous section is rigorous and complete. 
However, it fails to predict the behaviour actually observed in the laboratory when sectors are bent or unbent too severely: their compressed side does buckle, but earlier than predicted by the incremental theory, and  creases develop instead of smooth sinusoidal wrinkles, see \ref{leaflet} and \ref{silicone}.

This observation is well known and documented for the buckling of homogeneous solids, see  for example experimental pictures for the bending of blocks \cite{GeCh99, Suo09, RoGB10,DeMM10,Carf17}, of a cylinder \cite{Maha11} and of a sector \cite{d-o-m}, the torsion of a cylinder \cite{CiDe14} and of a tube \cite{Carf17}, the eversion of a tube \cite{Lian16,Carf17} and the shear-box deformation of a block \cite{Carf17}.
It has also been successfully captured by Finite Element (FE) simulations, see the seminal articles by Hong et al. \cite{Suo09}, Hohlfeld and Mahadevan \cite{Maha11}, and Cao and Hutchinson \cite{CaHu12, CaHu12b} (the latter include a nonlinear post-bifurcation analysis and imperfection sensitivity).
Note that there are very few FE simulations of creases in cylindrical coordinates \cite{BaKC14,Lian16,CDGT16}.

For our table-top experiments, we prepared a circular sector (Sector \fbox{$D$}) of soft silicone of height 58 mm, inner radius $A = 35$ mm, outer radius $B = 72$ mm (so that $\rho = A/B = 0.486$), and reference angle $\alpha_r = \pi/3$. 
We used superglue to attach two $70\times 70$ mm$^2$ squares of acrylic glass to the end faces of the sectors to bend or unbend the sector by applying torques mostly, and as little normal forces as possible.
Here the sector is not stretched axially ($\lambda_z=1$) and the scaled circumferential stretch $\lambda$ defined in \ref{not} coincides with $\lambda_\theta$ of \ref{stretches}.  
According to the incremental analysis summarised in \ref{fig_sim_3}, for this sector we should expect nine wrinkles to form in bending when $\alpha_d=0.9124 \pi$,  and six in eversion when $\alpha_d=-0.4372 \pi$. 
We collected the results of the incremental stability analysis for Sector \fbox{$D$} on the last line of \ref{tab:Table_1}.

In practice, we do not observe the formation of sinusoidal wrinkles in bending, but instead the formation of about eight creases. 
Moreover, although these creases are regularly spaced, some of them sometimes merge (period-doubling), depending on a given bending event.
In unbending the surface of the sector does not buckle.
In eversion, it buckles with a single deep crease in the middle of the everted sector, see \ref{silicone}.
However, we note that a perfect unbending so that the everted block has a circular shape is very hard to effect in practice. 

To investigate numerically the formation of wrinkles/creases, we implemented  FE models in ABAQUS/Explicit. 
For computational efficiency, we considered a 2D sector only, as in any case we are primarily interested in prismatic buckling. 
We chose a long time for the analysis to ensure a quasi-static deformation. 
Because perfect incompressibility is not possible in ABAQUS/Explicit, we used the neo-Hookean model with an initial bulk modulus 100 larger than the initial shear modulus to achieve near-incompressibility. 
We used linear reduced integration quadrilateral elements (CPE4R).

When a displacement is prescribed (say of a side of the sector from one location to another), ABAQUS implements it as taking place along a straight line. Thus, if one applies a displacement which deforms an undeformed sector into a closed tube (through bending or unbending), the intermediate deformation is not in agreement with the deformation described by \ref{deformation1}, \ref{radii} for deforming an undeformed sector into a bent or unbent sector. To solve this problem, we thus implemented a sequence of small displacements rather than a single large displacement.
Here, to ensure the sector was deformed along the `correct' path, that is the path closest to that of the exact solution given by \ref{deformation1}, \ref{radii}, we considered $N$ deformations:
 \begin{equation}
r=r(R),\qquad \theta=i \frac{\kappa}{N}\Theta, \qquad z=\lambda_z Z,
\end{equation}
for $i=1,2,3, \ldots, N$, where $\kappa=\pi/\alpha_r$ for bending and $\kappa=-\pi/\alpha_r$ for unbending.  
Here, the $N$-th deformation is the deformation which closes the sector completely, while the deformation for $i<N$ refers to an intermediate deformation.
For each of these deformations, we calculated the deformed geometry at each node of the two end faces and the non-buckling face using \ref{radii}, \ref{chib} and \ref{sigma1}. From here, we calculated the displacements necessary to go from the $i$-th deformation to the $(i+1)$-th deformation, resulting in a set of $N$ displacements for each node.
Then we created $N$ steps in ABAQUS, in each step imposing the calculated displacements at each node of the two end faces and on the non-buckling face. In practice, we used about $100$ steps. 
Zero tractions boundary condition was used on the remaining surface where the buckling will occur.

To initiate the buckling of the contracted face, we added a sinusoidal geometry perturbation of very small amplitude along that contour \cite{CaHu12, BaKC14}. 
Effectively, the amplitude was three orders of magnitude smaller than the radii. 

The simulations revealed the spontaneous formation of creases, with no smooth transition from the sinusoidal perturbation. 
The creases deepen quickly as the deformation progresses, and their sides come into contact, consistent with the analysis for compression of a half-space \cite{CaHu12}. 
There was also a spontaneous merging of some adjacent creases to form period-doubling patterns, so that the number of final creases was often less than the number $n$ of wrinkles that they emerged from. 

\begin{figure}[h]
\begin{center}
\includegraphics[width=1.0\linewidth]{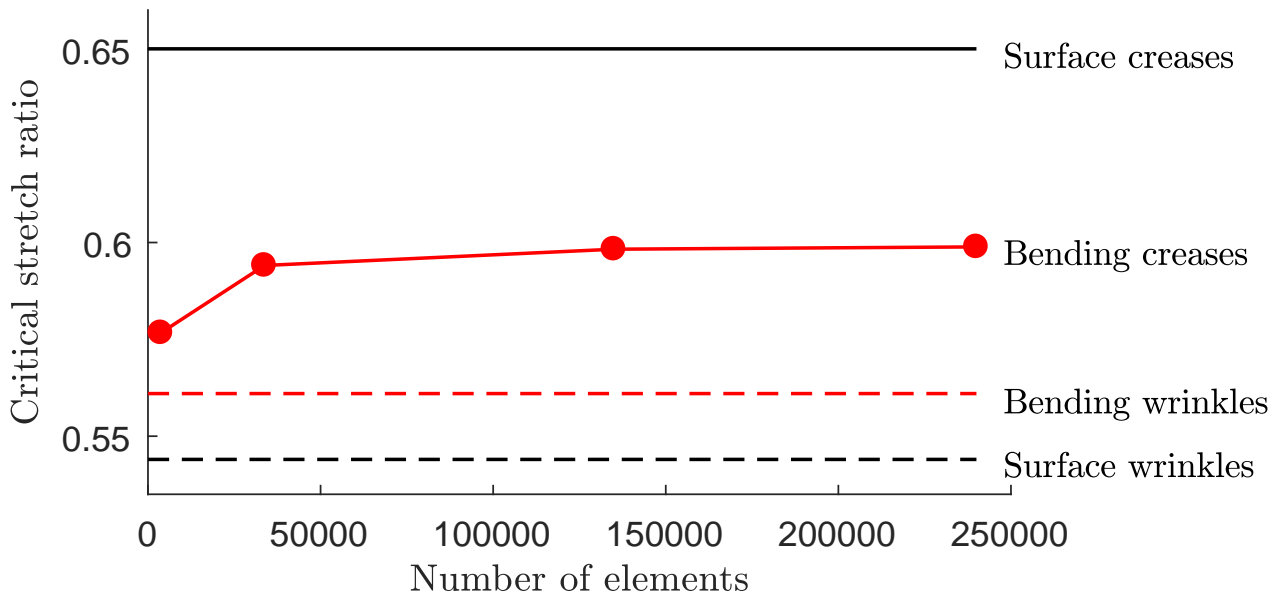}
\end{center}
\caption{Red point line: convergence with the refinement of the mesh of the critical circumferential stretch of compression for the formation of creases, here when bending Sector \fbox{$C$}. For comparison, the level of critical stretch is given for wrinkles (red dashed line below). We see that creases occur at about 4\% compression earlier than wrinkles. In contrast, creases appear on the surface of a half-space compressed in plane strain more than 10\% compression earlier (upper black full line)  than wrinkles (lower dashed black line) \cite{Suo09}.}
\label{mesh}
\end{figure}

We conducted a mesh sensitivity analysis with respect to the onset of buckling, which was identified by a drop in the elastic energy per unit thickness of the creased sector compared to that in the smooth body \cite{Suo09}.
We generally found that buckling occurred earlier every time the mesh was made finer. 
The solution started to converge as the number of elements increased, but eventually a very large number of elements was required to ascertain the limit, up to 240,000 elements, see \ref{mesh}. 
We had to use a gradient of mesh refinement near the compressed face and to book time on a high-end computer to perform these calculations.  
An alternative would have been to add a thin coating of vanishing stiffness \cite{Maha11}.

\begin{figure}[h]
\begin{center}
\includegraphics[width=1.0\linewidth]{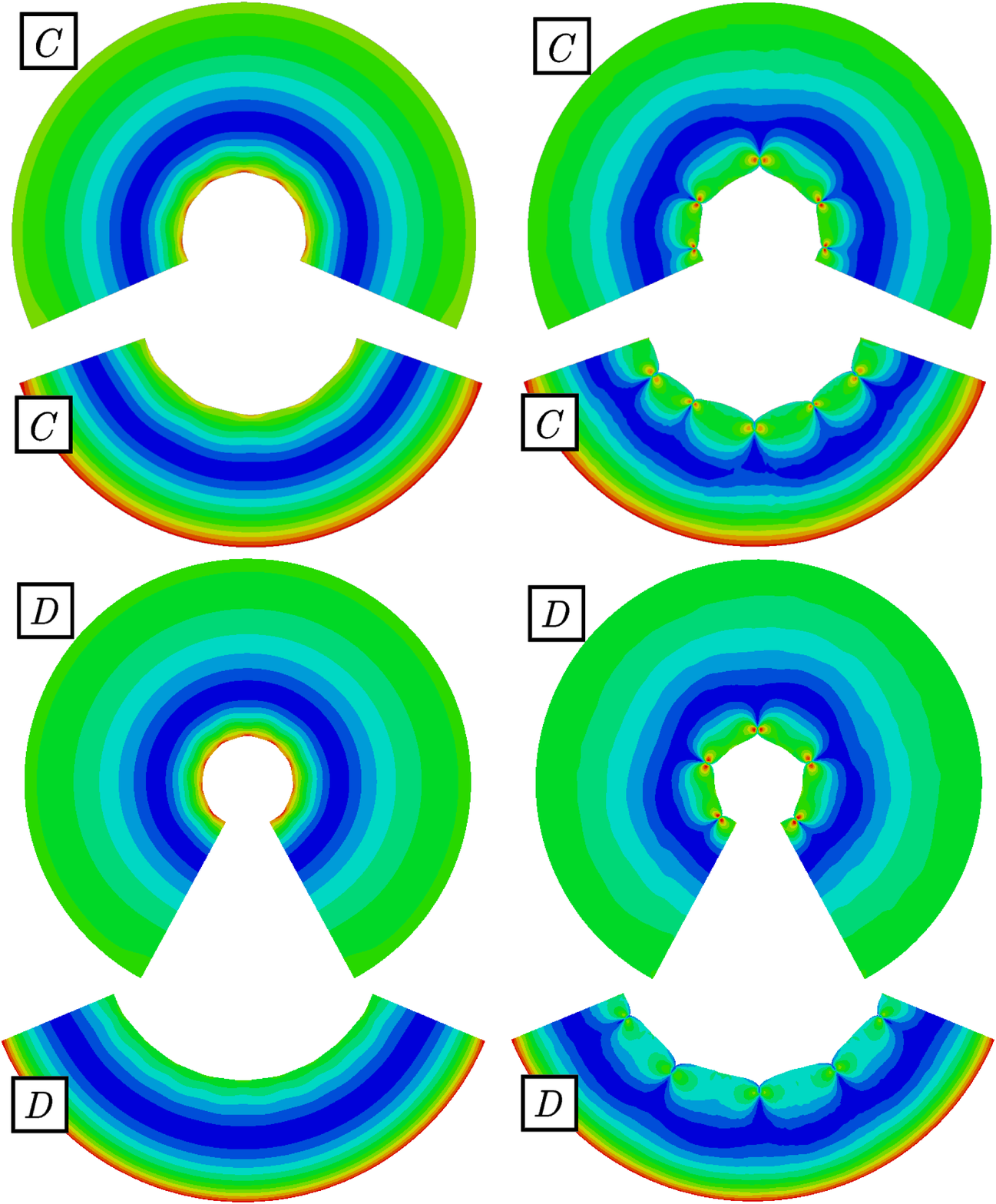}
\end{center}
\caption{Finite Element solutions of bending and unbending immediately before and after buckling. From top to bottom: bending sector \fbox{$C$},  unbending sector \fbox{$C$},  bending sector \fbox{$D$}, unbending sector \fbox{$D$}. The colours correspond to the von Mises stress level, from blue (low) to red (high).}
\label{abaqus}
\end{figure}

We also conducted a spectral analysis with respect to the linear sinusoidal perturbation, that is, we performed the simulations for $m=2,3,4,\ldots$ wrinkles, and kept the one for which the buckling occurred the earliest. 
For most cases (with the exception of sector \fbox{$D$} in unbending), the number of periods in the sinusoidal perturbation corresponding to earliest buckling was the same or close to the number of wrinkles predicted by the linear analysis. We also note that the number $m$ is evident through the development of stress concentrations along the buckling faces in the lead up to buckling, see the images on the left-hand side of  \ref{abaqus}.
We conclude that the incremental analysis tends to provide a good indication of the optimal shape for the perturbation.

For the simulation of the physical Sector \fbox{$D$} we used 183,750 elements. We found that buckling occurred at its earliest in bending when $m=9$ for the deformed angle $\alpha_d=0.842 \pi$, and $m= 9$ in unbending when $\alpha_d=-0.374 \pi$, with period-doubling occurring at several locations, see \ref{abaqus}.

For Sector  \fbox{$C$}, we used 240,000 elements.
We found that buckling corresponded to $m=7$ in bending when $\alpha_d=0.633 \pi$, and to $m=5$ in unbending when $\alpha_d=-0.392 \pi$, with period-doubling occurring again.

The crease formations which occurred in our finite element simulations are  consistent with our table-top experiments (at least with the bending experiment; the eversion experiment with its single crease is not well captured by the modelling, which assumed a circular everted sector, a geometry which is impossible to obtain in practice). 
But there are differences with the non-linear stability analyses of crease formation conducted previously.

For example, Hong et al. \cite{Suo09} showed that a semi-infinite body of neo-Hookean material creases in plane strain  at a critical amount of stretch equal to 0.65, which is 11\% strain earlier than 0.54, the critical stretch for wrinkles found by Biot \cite{Biot63}. 
Here, we found that for our sectors the difference between crease onset and wrinkle onset was 4\% or less. 
Also, half-space crease analysis does not provide a wavelength for the crease, since there is no characteristic length in that context. 
Here we note the agreement (or near agreement) between the number of creases in the finite element simulations and the number of wrinkles predicted by the linear analysis.

In conclusion, the incremental stability provides valuable information on the loss of stability for the large bending or unbending of a  circular sector. 
It will also be quite straightforward to extend it to material models other than Mooney-Rivlin.
Finally, it provides the basis for the study of coated materials where  sinusoidal wrinkles are expected to dominate \cite{CaHu12}.

\section*{Acknowledgements}

MD and RM gratefully acknowledge the financial support of the Irish Research Council. LS and TS are grateful for the support received from the Schulich School of Engineering and Zymetrix Biomaterials \& Tissue Engineering Technology Development Centre, University of Calgary. LV would like to thank the Carnegie Trust for the financial support (R\&E Project Code: 67954/1). 
This publication has also been made possible by a James M Flaherty Research Scholarship from the Ireland Canada University Foundation, with the assistance of the Government of Canada/avec l'appui du gouvernement du Canada.

The authors gratefully acknowledge the SFI/HEA Irish Centre for High-End Computing (ICHEC) for the provision of computational facilities and support. 
%\disclaimer{Insert disclaimer text here.}

\section*{Appendix A: Proofs of existence and uniqueness; Thin-wall expansions}

%%%%%%%%%%%%%%%%%%%%%%%%%%%
\setcounter{section}{1}

\renewcommand{\theequation}{\Alph{section}.\arabic{equation}}

%-==========================

\subsection{Existence and uniqueness}

%===========================

We assume that the strain-energy function $W$ satisfies the \emph{strong ellipticity condition}. 
As shown by Ogden \cite{ray}, it amounts to
\begin{equation}\label{strong}
\dfrac{\lambda}{\lambda^2-1}\widehat{W}'(\lambda)>0,\qquad 
\lambda^2\widehat{W}'' + \dfrac{2\lambda}{\lambda^2+1}\widehat{W}'(\lambda) > 0.
\end{equation}
From these inequalities, respectively, we deduce that
\begin{equation}\label{wprime}
\widehat{W}'(\lambda)\gtreqqless0 \quad(\textrm{as} \:\:\lambda\gtreqqless 1)\quad\textrm{and}\quad\widehat{W}''(1)>0.
\end{equation}
Finally,  by integrating the second inequality (\ref{strong}$_2$)  we deduce that, in a right neighbourhood of $0$,
\begin{equation}
\frac{\lambda^2}{\lambda^2 + 1}|\widehat{W}'(\lambda)|>c,
\end{equation}
for some positive constant $c$, from which follows that
\begin{equation}\label{int0}
\lim_{\lambda\rightarrow0^+}\lambda^2|\widehat{W}'(\lambda)|\geq c>0.
\end{equation}

We now investigate the existence and uniqueness of a positive root to \ref{sigma1}$_2$, which we recall here:
\begin{equation} \label{lambda_ab}
\int_{\lambda_{a}}^{\lambda_b}\frac{\widehat{W}'(s)}{\kappa-s^2} \text d s=0.
\end{equation} 
 
First, assume that $\kappa\in[-2\pi/\alpha_r,0[$. 
Then $\lambda_a^2>(1-\rho^2)|\kappa|/\rho^2$ and $\lambda\in[\lambda_b,\lambda_a]$. 
It follows that if $\lambda_b\geq1$ or $\lambda_a\leq1$, then the integrand has the same sign over the entire range of integration and hence \ref{lambda_ab} does not admit a solution. 
Hence we must have $\lambda_b<1<\lambda_a$, or, using \ref{chib}, 
\begin{equation}\label{range-}
\frac{\sqrt{(1-\rho^2)|\kappa|}}{\rho}<\lambda_a<\frac{\sqrt{1+(1-\rho^2)|\kappa|}}{\rho}.
\end{equation}
To prove the existence of a root for \ref{lambda_ab} we set 
\begin{equation}
\lambda^*=\max\left\{\frac{\sqrt{(1-\rho^2)|\kappa|}}{\rho},1\right\},
\end{equation}
and define the function $f$ as
\begin{equation}
f:y\in\left[\lambda^*,\frac{\sqrt{1+(1-\rho^2)|\kappa|}}{\rho}\right]\mapsto  \int_{\sqrt{\rho^2y^2+(1-\rho^2)\kappa}}^y\frac{\widehat{W}'(s)}{\kappa-s^2}\text d s.
\end{equation}

Next, since $\sqrt{\rho^2y^2+(1-\rho^2)\kappa}<1<y$ for all $y\in\mathrm{Dom}(f)$ and $\kappa-y^2$ is negative,  from \ref{strong}$_1$  we conclude that
\begin{equation}
f'(y)=\frac{1}{\kappa-y^2}\left[\widehat{W}'(y)-\widehat{W}'\left(\displaystyle\sqrt{\rho^2y^2+(1-\rho^2)\kappa}\right)\frac{y}{\sqrt{\rho^2y^2+(1-\rho^2)\kappa}}\right]<0,
\end{equation}
whence $f$ is a decreasing function.
For $|\kappa|\leq \rho^2/(1-\rho^2)$, 
%We now observe that
%\begin{equation}
%\lambda^*=\left\{\begin{array}{ll}
%1 \quad \textrm{if } &|\kappa|\leq \frac{\rho^2}{1-\rho^2},\\
%[3mm]
%\frac{\sqrt{(1-\rho^2)|\kappa|}}{\rho} \quad \textrm{if } &|\kappa|\leq \frac{\rho^2}{1-\rho^2},
%\end{array}\right.
%\end{equation}
from \ref{wprime}$_1$ we deduce that
\begin{equation}
f(\lambda^*)=f(1)=\int_{\sqrt{\rho^2+(1-\rho^2)\kappa}}^{1}\frac{\widehat{W}'(s)}{\kappa-s^2}\text d s>0,
\end{equation}
and for $|\kappa|>\rho^2/(1-\rho^2)$,  from \ref{int0} we deduce that
\begin{equation}\label{lambstar}
f(\lambda^*)=f\left(\frac{\sqrt{(1-\rho^2)|\kappa|}}{\rho}\right)=\int_{0}^{\frac{\sqrt{(1-\rho^2)|\kappa|}}{\rho}}\frac{\widehat{W}'(s)}{\kappa-s^2}\text d s=+\infty.
\end{equation}
At the other end of the interval where $f$ is defined we have, as a consequence of  \ref{wprime}$_1$,
\begin{equation}
f\left(\frac{\sqrt{1+(1-\rho^2)|\kappa|}}{\rho}\right)=\int_{1}^{\frac{\sqrt{1+(1-\rho^2)|\kappa|}}{\rho}}\frac{\widehat{W}'(\lambda)}{\kappa-s^2}\text d s<0.
\end{equation}
In conclusion, $f$ has exactly one zero in its domain and so, when $\kappa\in[-2\pi/\alpha_r,0[$,  \ref{lambda_ab} admits a unique root in the range $\left]\lambda^*,{\sqrt{1+(1-\rho^2)|\kappa|}}/{\rho}\right[$.

We assume now that $0<\kappa<1$. 
If $0<\lambda_a<\sqrt{\kappa}$,  then  the integrand in  \ref{lambda_ab} is strictly negative and the equation does not admit a solution. 
Thus we must have $\sqrt{\kappa}<\lambda_b<1<\lambda_a$, or using \ref{chib},
\begin{equation}
\kappa<1<\lambda_a<\sqrt{1-(1-\rho^2)\kappa}/\rho.
\end{equation}
Consider the function $g$ defined as
\begin{equation}
g:y\in\left[1,\frac{\sqrt{1-(1-\rho^2)\kappa}}{\rho}\right]\mapsto \int_{\sqrt{\rho^2y^2+(1-\rho^2)\kappa}}^y\frac{\widehat{W}'(s)}{\kappa-s^2}\text d s.
\end{equation}
With the aid of  \ref{wprime}$_1$, we deduce in turn that
\begin{gather}
g'(y) =  \frac{1}{\kappa-y^2}\left[\widehat{W}'(y)-\widehat{W}'\left(\sqrt{\rho^2y^2+(1-\rho^2)\kappa}\right)\frac{y}{\sqrt{\rho^2y^2+(1-\rho^2)\kappa}}\right]<0,\\g\left(\frac{\sqrt{1-(1-\rho^2)\kappa}}{\rho}\right)=\int_{1}^{\frac{\sqrt{1-(1-\rho^2)\kappa}}{\rho}}\frac{\widehat{W}'(s)}{\kappa-s^2}\text d s<0,\quad g(1)=\int_{\sqrt{\rho^2+(1-\rho^2)\kappa}}^{1}\frac{\widehat{W}'(s)}{\kappa-s^2}\text d s>0.
\end{gather}
From these we conclude that $g$ has exactly one zero in its domain and so \ref{lambda_ab} admits a unique solution 
\begin{equation}\label{k<1}
\lambda_a\in\left]1,\frac{\sqrt{1-(1-\rho^2)\kappa}}{\rho}\right[.
\end{equation}

Finally in the case $\kappa>1$ it is easy to show that \ref{lambda_ab} admits a solution only if $\lambda_a<1<\lambda_b<\sqrt{\kappa}$, or using \ref{chib},
\begin{equation}
\lambda^{\star}<\lambda_a<1,
\end{equation}
where
\begin{equation}
\lambda^{\star}=\sqrt{\max\left\{0,1-(1-\rho^2)\kappa/\rho^2\right\}}.
\end{equation}

Then we  consider the function $h$ defined as 
\begin{equation}
\displaystyle h:y\in\left[\lambda^{\star},1\right]\mapsto \int_y^{\sqrt{\rho^2y^2+(1-\rho^2)\kappa}}\frac{\widehat{W}'(s)}{\kappa-s^2}\text d s.
\end{equation}
Assume first  that $0<\kappa<1/(1-\rho^2)$. Then,
%\begin{equation}
%\max\{0,\chi_m[\eps(1-\kappa)+1]\}=\left\{\begin{array}{ll}
%0 &\quad \textrm{if } \kappa\geq \frac{\eps+1}{\eps},\\
%\chi_m[\eps(1-\kappa)+1] &\quad \textrm{if } 1<\kappa<\frac{\eps+1}{\eps}.
%\end{array}\right.
%\end{equation}
with the aid of  \ref{wprime}$_1$, we find that
\begin{equation}\label{h1}
h(\lambda^{\star})=h\left(\sqrt{1-(1-\rho^2)\kappa}/\rho\right)=\int_{\sqrt{1-(1-\rho^2)\kappa}/\rho}^{1}\frac{\widehat{W}'(s)}{\kappa-s^2}\text d s<0.
\end{equation}
Conversely, when $\kappa\geq 1/(1-\rho^2)$, we have $\lambda^{\star}=0$ and, in view  of \ref{int0}, we deduce that 
\begin{equation}\label{h2}
\lim_{y\rightarrow0^+}h(y)=\int_{0}^{\sqrt{(1-\rho^2)\kappa}}\frac{\widehat{W}'(s)}{\kappa-s^2}\text d s=-\infty.
\end{equation}
Thanks  to \ref{wprime}$_1$ we have 
\begin{equation}
\displaystyle h(1)=\int_{1}^{\sqrt{\rho^2+(1-\rho^2)\kappa}}\frac{\widehat{W}'(s)}{\kappa-s^2}\text d s>0,
\end{equation}
and
\begin{equation}\label{h4}
h'(y) = \dfrac{1}{\kappa-y^2}\left[\widehat{W}'\left(\sqrt{\rho^2y^2+(1-\rho^2)\kappa}\right)\frac{y}{\sqrt{\rho^2y^2+(1-\rho^2)\kappa}}-\widehat{W}'(y)\right]>0.
\end{equation}
It follows that $h$ has exactly one zero in its domain and thus \ref{lambda_ab} admits a unique solution 
\begin{equation}\label{k>1}
\lambda_a\in\left]\sqrt{\max\left\{0,\frac{1-(1-\rho^2)\kappa}{\rho^2}\right\}},1\right[.
\end{equation}

Finally, it is worth noting that  from \ref{k<1} and \ref{k>1}  $\lambda_a\rightarrow 1$ as $\kappa\rightarrow1$. On the other hand, from \ref{chib} and \ref{strong}$_1$ we readily deduce that the unique root to \ref{lambda_ab} with $\lambda_a=1$ is   $\kappa=1$.  In other words, the axial stretching (\ref{homogeneous}) with $\kappa=1$ is the only admissible homogeneous deformation.

%======================

\subsection{Thin-walled sectors}

%======================

For thin sectors, we perform an asymptotic analysis in the small thickness parameter $\varepsilon>0$ defined as:
$
\varepsilon= 1-\rho\ll1.
$

First we rewrite the left-hand side of \ref{lambda_ab} as a function $F$ of $\varepsilon$, specifically
\begin{equation}
F(\varepsilon)=\int_{\lambda_a}^{\sqrt{(1-\varepsilon)^2\lambda_a^2+ \varepsilon(2 - \varepsilon)\kappa}}\dfrac{\widehat{W}'(\lambda)}{\kappa-\lambda^2}\mathrm{d}\lambda.\label{new f}
\end{equation}
Expanding $F(\varepsilon)$ as a Maclaurin series in $\varepsilon$ up to the fifth order, substituting into the equation $f(\lambda_b)=0$ and dropping a common factor $\varepsilon$, yields the equation
\begin{equation}
\label{expansion1}
\widehat{W}'\left(\lambda_a\right)+F^{(1)}\varepsilon+F^{(2)}\varepsilon^2+F^{(3)}\varepsilon^3+F^{(4)}\varepsilon^4+ \mathcal O(\varepsilon^5)=0,
\end{equation}
where
\begin{align}
 F^{(1)}&=\frac{1}{\lambda_a^2}\left[(2\lambda_a^2-\kappa)\widehat{W}'(\lambda_a)-\lambda_a(\lambda_a^2-\kappa)\widehat{W}''\left(\lambda_a\right)\right],
\notag \\[6pt]
 F^{(2)}&= \frac{1}{\lambda_a^4}\left[(6\lambda_a^4-7\lambda_a^2\kappa+3\kappa^2)\widehat{W}'(\lambda_a)\right.
\notag \\
&  \left. \qquad   -\lambda_a(4\lambda_a^4-7\lambda_a^2\kappa+3\kappa^2)\widehat{W}''(\lambda_a) 
+ \lambda_a^2(\lambda_a^2-\kappa)^2\widehat{W}'''(\lambda_a)\right],\notag \\[6pt]
 F^{(3)}&= \frac{1}{24\lambda_a^6}\Big[3(8\lambda_a^6-16\lambda_a^4\kappa+15\lambda_a^2\kappa^2-5\kappa^3)\widehat{W}'(\lambda_a)
\notag \\
&   \qquad -3\lambda_a(6\lambda_a^6-16\lambda_a^4\kappa+15\lambda_a^2\kappa^2-5\kappa^3)\widehat{W}''(\lambda_a)+\\
&  \qquad\qquad +6\lambda_a^2(\lambda_a^2-\kappa)^3\widehat{W}'''(\lambda_a)-\lambda_a^3(\lambda_a^2-\kappa)^3\widehat{W}^{(iv)}(\lambda_a)\Big],
\notag \\[6pt]
F^{(4)}&= \frac{1}{120\lambda_a^8}\Big[3(40\lambda_a^8-120\lambda_a^6\kappa+183\lambda_a^4\kappa^2-130\lambda_a^2\kappa^3+35\kappa^4)\widehat{W}'(\lambda_a)
\notag \\
&\qquad \qquad -\lambda_a(96\lambda_a^8-360\lambda_a^6\kappa  + 539\lambda_a^4\kappa^2-390\lambda_a^2\kappa^3 + 105\kappa^4)\widehat{W}''(\lambda_a) \notag \\ 
&\qquad \qquad\qquad + 3\lambda_a^2(12\lambda_a^8-50\lambda_a^6\kappa + 79\lambda_a^4\kappa^2 - 56\lambda_a^2\kappa^3 + 9\kappa^4)\widehat{W}'''(\lambda_a) 
 \notag \\
 &\qquad\qquad \qquad\qquad -2\lambda_a^3(4\lambda_a^8-17\lambda_a^6\kappa+27\lambda_a^4\kappa^2-19\lambda_a^2\kappa^3+5\kappa^4)\widehat{W}^{(iv)}(\lambda_a)
 \notag \\
 & \qquad\qquad\qquad\qquad\qquad
 + \lambda_a^4(\lambda_a^8-4\lambda_a^6\kappa+6\lambda_a^4\kappa^2-4\lambda_a^2\kappa^3+\kappa^4)\widehat{W}^{(v)}(\lambda_a)\Big].
\end{align}

Next, we expand $\lambda_a$ in terms of $\varepsilon$ to the fourth order,
\begin{equation}
\lambda_a=\lambda_a^{(0)}+\lambda_a^{(1)}\varepsilon+\lambda_a^{(2)}\varepsilon^2+\lambda_a^{(3)}\varepsilon^3+\lambda_a^{(4)}\varepsilon^4 + \mathcal O(\varepsilon^5),\label{lambdaaexpansion}
\end{equation}
where the $\lambda^{(i)}$ are determined in turn as follows.

Substituting the expansion of  $\lambda_a$ into the previous expansion (\ref{expansion1}) and equating to zero the coefficients of each power in the resulting expression, we obtain first, at zero order, that
$
\widehat{W}'\left(\lambda_a^{(0)}\right)=0,
$
and hence, by \ref{wprime}$_2$, that $\lambda_a^{(0)}=1$.

Using this result in the first-order term, we  then obtain
\begin{equation}
\left[\frac{1}{2}(\kappa-1)+\lambda_a^{(1)}\right]\widehat{W}''\left(1\right)=0,
\end{equation}
and because $\widehat{W}''\left(1\right)>0$, we deduce that $\lambda_a^{(1)}=(1-\kappa)/2$.  

Then, the second-order term yields
\begin{equation}
\left[\lambda_a^{(2)}+\frac{5}{12}(\kappa-1)\right]\widehat{W}''\left(1\right)+\frac{(\kappa-1)^2}{24}\widehat{W}'''\left(1\right)=0.
\end{equation}
The resulting expression for $\lambda_a$, up to the second order in $\varepsilon$, is therefore
\begin{equation}
\lambda_a=1+\frac{1}{2}(1-\kappa)\varepsilon+\frac{1-\kappa}{24}\left[10-\frac{\widehat{W}'''(1)}{\widehat{W}''(1)}(1-\kappa)\right]\varepsilon^2 + \mathcal O(\varepsilon^3).\label{lambdaaWW}
\end{equation}
However,  in virtue of the universal result: $\widehat{W}'''(1)/\widehat{W}''(1)=-3$ (see for example \cite{ray85}),  the above formula reduces to
\begin{equation} 
\lambda_a=1-\frac{1}{2}(1-\kappa)\varepsilon+\frac{1}{24}(1-\kappa)(13-3\kappa)\varepsilon^2 + \mathcal O(\varepsilon^3).
\end{equation}

Proceeding in a similar way (and omitting the lengthy details), we obtain,  up to the fourth order in $\varepsilon$,
\begin{align} \label{lambdab4order}
 \lambda_a & =  1+\frac{1}{2}(1-\kappa)\varepsilon 
 + \dfrac{1}{24}(1-\kappa)(13-3\kappa)\varepsilon^2-\frac{1}{48}(1-\kappa)(3\kappa^2+8\kappa-27)\varepsilon^3
 \notag \\
& \qquad  + \frac{1}{5760}(1-\kappa)\Big[45\kappa^3-363\kappa^2-1813\kappa+3667
\notag \\
&  \qquad \qquad + (1-\kappa)^2\frac{2(15\kappa-23)\widehat{W}^{(iv)}(1) - 3(1-\kappa)\widehat{W}^{(v)}(1)}{\widehat{W}''(1)}\Big]\varepsilon^4 
+ \mathcal O(\varepsilon^5).
\end{align}
Note, in particular, that the results are independent of the strain energy function up to order $\varepsilon^3$. 

\setcounter{section}{2}
\setcounter{equation}{0}

%%%%%%%%%%%%%%%%

\section*{Appendix B: Algorithms for the analysis of the Stroh problem}

%%%%%%%%%%%%%%%%

Here we outline two numerically robust methods to obtain the numerical solution of the Stroh problem (\ref{stroh}). 

The first one is called the \emph{compound matrix method}. 
In this method, we let $\boldsymbol{\eta}^{(1)}$, $\boldsymbol{\eta}^{(2)}$ be two linearly independent solutions of  \ref{stroh}, and use them
to generate the six compound functions $\phi_1=\left\langle \eta_1,\eta_2\right\rangle$, $\phi_2=\left\langle \eta_1,\eta_3\right\rangle$, $\phi_3=i\left\langle \eta_1,\eta_4\right\rangle$, $\phi_4=i\left\langle \eta_2,\eta_3\right\rangle$, $\phi_5=\left\langle \eta_2,\eta_4\right\rangle$, $\phi_6=\left\langle \eta_3,\eta_4\right\rangle$, where $\left\langle \eta_i,\eta_j\right\rangle \equiv \eta_i^{(1)}  \eta_j^{(2)}
 - \eta_i^{(2)} \eta_j^{(1)}$. 
 Now, computing the derivatives of $\phi_i$ with respect to $r$ yields the so-called compound equations 
\begin{equation}\label{compound equations}
\dfrac{\text d \boldsymbol{\phi}}{\text d r} = \frac{1}{r}\mathbf{A}(r)\boldsymbol{\phi}(r),
\end{equation}
where $ \boldsymbol{\phi}=(\phi_1,\dots,\phi_6)^\mathrm{T}$ and $\mathbf{A}$, the \emph{compound matrix}, has the form
\begin{equation}
\mathbf{A}=\left(\begin{array}{cccccc}
-\sigma_{rr}/\alpha & 0 &-1/\alpha & 0 &0 &0 \\
-\kappa_{12} & 0 & -n(1-\sigma_{rr}/\alpha) & -n & 0 & 0 \\
-\kappa_{22}  & n & -(2-\sigma_{rr}/\alpha) & 0 &n & 0\\
\kappa_{11} & n(1-\sigma_{rr}/\alpha) & 0 &(2-\sigma_{rr}/\alpha) &n(1-\sigma_{rr}/\alpha) & -1/\alpha\\
-\kappa_{12} & 0& -n(1-\sigma_{rr}/\alpha) & -n & 0 & 0\\
0 & -\kappa_{12} & \kappa_{11} & - \kappa_{22} & -\kappa_{12} & \sigma_{rr}/\alpha
\end{array}\right).\label{compound matrix}
\end{equation}
The compound equations (\ref{compound equations}) must be integrated numerically, starting with the initial condition
$
\boldsymbol{\phi}(a)=\phi_1(a)[1,0,0,0,0,0]^\mathrm{T},\label{init}
$
and aiming at the target condition
$
\phi_6(b)=0,\label{target}
$
according to \ref{bcstroh}.
In passing, note that $\mathbf A$ is clearly singular, as in the straightening problem (noticed by \cite{DOSV14a,DOSV14b}) and in the bending of a straight block (unnoticed \cite{Haug99,CoDe08,DeAC09,RoGB10}).
However, it turns out that the singularity of the matrix does not affect the efficiency of the integration scheme.

The second approach is called the \emph{surface impedance matrix method}. 
In this method, we  define the matricant solution matrix $\boldsymbol{M}(r,r_c)=\begin{pmatrix}\boldsymbol{M}_1(r,r_c)&\boldsymbol{M}_2(r,r_c)\\\boldsymbol{M}_3(r,r_c)&\boldsymbol{M}_4(r,r_c)\end{pmatrix}$ to \ref{stroh} such as $\boldsymbol{\eta}(r)=\boldsymbol{M}(r,r_c)\boldsymbol{\eta}(r_c)$
(clearly $\boldsymbol{M}(r_c,r_c)$ is  the identity matrix).
Here $r_c$ can be either $r_a$ or $r_b$, depending on what is most convenient. 
This allows us to write the initial boundary conditions in the simple form $\boldsymbol{z}^c(r_c)= \mathbf 0$, where $\boldsymbol{z}^c=-i\boldsymbol{M}_{3}(r,r_c)\boldsymbol{M}_{1}(r,r_c)^{-1}$ is called the conditional impedance matrix. 
Now from the Stroh formalism (\ref{stroh}) we can derive two relevant equations
\begin{equation}
\label{ric}
\dfrac{\text d}{\text dr} \mathbf z^c= \displaystyle\frac{1}{r}\left[ \mathbf z^c\mathbf G_2 \mathbf z^c + \text i \left(\mathbf G_1\right)^\dagger \mathbf z^c -  \text i \mathbf z^c\mathbf G_1+ \mathbf G_3\right],\quad\dfrac{\text d}{\text dr} \mathbf U= \displaystyle\frac{1}{r}\left[  \text i \mathbf G_1 \mathbf U -  \mathbf G_2\mathbf z^c\mathbf U \right],\end{equation}
where $\mathbf G_i$ ($i=1,2,3$) are subblocks of matrix $\mathbf G$ from \ref{stroh}.
The numerical integration of  \ref{ric}$_1$, a differential Riccati equation,  for $r_c=r_a$ with initial condition $\boldsymbol{z}^a(r_a)= \mathbf 0$, allows us to find the critical eigenvalues, i.e. critical deformation angles and stretches, of the Stroh problem (\ref{stroh}) upon satisfaction of the boundary condition on the other face of the sector, which is $\det\boldsymbol{z}^a(r_b)= \mathbf 0$ (the latter one is equivalent to $V(r_b)/U(r_b)=-z^a_{11}(r_b)/z^a_{12}(r_b)=-z^a_{21}(r_b)/z^a_{22}(r_b)$.) 
Next, the corresponding eigenvectors of the Stroh problem (\ref{stroh}) are obtained through the simultaneous numerical integration of the two \ref{ric} for $r_c=r_b$, with initial condition $\boldsymbol{z}^b(r_b)= \mathbf 0$ and $\mathbf U(r_b)=U(r_b)[1,-z^a_{11}(r_b)/z^a_{12}(r_b)]^\mathrm{T}$.

\ref{tab:Table_3} gives a detailed numerical algorithm to solve the impedance and compound matrix equations. 
\begin{table}[!h]
\vspace{-1em}
\caption{Numerical implementation of the impedance and compound matrix methods.}%%%Table caption goes here
\label{table_example}
\begin{tabular}{l}%%%The number of columns has to be defined here
\hline
Define reference geometry, e.g. $\alpha_r$ and $\rho$\vspace{-0.1em}\\
DO For different mode numbers m=1,2,3... \\
\quad\quad DO For all deformations $\kappa<\pi/\alpha_r$ in bending (or $\kappa>-\pi/\alpha_r$ in unbending and eversion)\vspace{0.3em}\\
\quad\quad\quad\quad$n=m\pi/(\kappa\alpha_r)$;\vspace{0.3em}\\
\quad\quad\quad\quad Find $\lambda_a$ and $\lambda_b$;\vspace{0.5em}\\
**************************** In case of the compound matrix method approach***************************\\
\quad\quad\quad\quad\vline\;\vline Integrate the compound matrical differential equation $\displaystyle\frac{\text d \boldsymbol{\phi}}{\text d r}=\frac{1}{r}\mathbf{A}\boldsymbol{\phi}$ on $r\in(a,b)$ with\vspace{-0.0em}\\
 \quad\quad\quad\quad \vline\;\vline the boundary condition $\boldsymbol{\phi}(a)=\phi_1(a)[1,0,0,0,0,0]^\mathrm{T}$; \vspace{0.3em}\\
 \quad\quad\quad\quad IF $\phi_6(b)=0$ (OR $\phi_6(b)$ is monotonic function of $\kappa$ and changes its sign) THEN\vspace{0.3em}\\
 \quad\quad\quad\quad\quad\quad Obtain critical values of $\kappa$, $\alpha_d$ and $\lambda_a$ in bending (or $\lambda_b$ in unbending and eversion);\vspace{0.3em}\\
 \quad\quad\quad\quad\quad\quad BREAK\vspace{0.3em}\\
 \quad\quad\quad\quad END IF\vspace{0.3em}\\
**************************** In case of the impedance matrix method approach***************************\\
\quad\quad\quad\quad \vline\;\vline Integrate the Riccati equation $
\dfrac{\text d}{\text dr} \mathbf z^a= \displaystyle\frac{1}{r}\left[ \mathbf z^a\mathbf G_2 \mathbf z^a + \text i \left(\mathbf G_1\right)^\dagger \mathbf z^a -  \text i \mathbf z^a\mathbf G_1+ \mathbf G_3\right]$ on $r\in(a,b)$\vspace{-0.1em}\\
\quad\quad\quad\quad \vline\;\vline with the boundary condition $\mathbf z^a(a)=\begin{pmatrix} 
0 & 0 \\
0 & 0 
\end{pmatrix}$;\vspace{0.3em}\\
 \quad\quad\quad\quad IF $\det\mathbf z^a(b)=0$ (OR $\det\mathbf z^a(b)$ is monotonic function of $\kappa$ and start to plummet) THEN\vspace{0.3em}\\
 \quad\quad\quad\quad\quad\quad Obtain critical values of $\kappa$, $\alpha_d$ and $\lambda_a$ in bending (or $\lambda_b$ in unbending and eversion);\vspace{0.3em}\\
 \quad\quad\quad\quad\quad\quad \vline\;\vline Integrate the Riccati equation $
\dfrac{\text d}{\text dr} \mathbf z^b= \displaystyle\frac{1}{r}\left[ \mathbf z^b\mathbf G_2 \mathbf z^b + \text i \left(\mathbf G_1\right)^\dagger \mathbf z^b -  \text i \mathbf z^b\mathbf G_1+ \mathbf G_3\right]$ together \vspace{-0.1em}\\
 \quad\quad\quad\quad\quad\quad \vline\;\vline with the $\dfrac{\text d}{\text dr} \mathbf U= \displaystyle\frac{1}{r}\left[  \text i \mathbf G_1 \mathbf U -  \mathbf G_2\mathbf U \right]$ on $r\in(b,a)$ using boundary conditions \vspace{-0.1em}\\
 \quad\quad\quad\quad\quad\quad \vline\;\vline$\mathbf z^b(b)=\begin{pmatrix}
0 & 0 \\
0 & 0 
\end{pmatrix}$ and $\mathbf U(b)=U(b)[1,-z^a_{11}(b)/z^a_{12}(b)]^\mathrm{T}$ to obtain the mechanical    \vspace{-0.1em}\\
 \quad\quad\quad\quad\quad\quad \vline\;\vline displacement field across the thickness of a given sector;\vspace{0.3em}\\
 \quad\quad\quad\quad\quad\quad BREAK\vspace{0.3em}\\
 \quad\quad\quad\quad END IF\vspace{0.3em}\\
\quad\quad END DO\vspace{0.3em}\\
END DO\vspace{0.3em}\\
\vline\;\vline  Determine the acute mode number among all considered modes $m=1,2,3...$, for which critical stret-\vspace{-0.1em}\\ \vline\;\vline ches $\lambda_a$ (bending) or $\lambda_b$ (unbending and eversion) are the highest. This allows to predict for a given\vspace{-0.1em}\\\vline\;\vline sector  when the
 buckling  will occur and in how many wrinkles it will result. \vspace{0.3em}\\
\hline
\end{tabular}
\label{tab:Table_3}
\end{table}%%%End of the table

%%%%%%%%%% Insert bibliography here %%%%%%%%%%%%%%

\end{document}